\documentclass[twocolumn,aps,prb,widetext,showpacs,superscriptaddress]{revtex4}
\usepackage{mathrsfs,bm,multirow,amsmath,amsfonts,amssymb,array,booktabs,float}
\usepackage{graphicx,times,graphics,color,epsfig}

\begin{document}

\title{Quantum Transport Theory with the Nonequilibrium Coherent Potentials}
\author{Yu Zhu}
\email{eric@nanoacademic.ca}
\affiliation{NanoAcademic Technologies Inc., 7005 Blvd. Taschereau, Brossard, QC, J4Z 1A7
Canada}
\author{Lei Liu}
\affiliation{NanoAcademic Technologies Inc., 7005 Blvd. Taschereau, Brossard, QC, J4Z 1A7
Canada}
\author{Hong Guo}
\affiliation{Department of Physics, McGill University, Montreal, QC, H3A 2T8, Canada}
\affiliation{NanoAcademic Technologies Inc., 7005 Blvd. Taschereau, Brossard, QC, J4Z 1A7
Canada}
\date{\today}

\begin{abstract}
Since any realistic electronic device has some degree of disorder,
predicting disorder effects in quantum transport is a critical problem. Here
we report the theory of nonequilibrium coherent potential approximation
(NECPA) for analyzing disorder effects in nonequilibrium quantum transport
of nanoelectronic devices. The NECPA is formulated by contour ordered
nonequilibrium Green's function where the disorder average is carried
out within the coherent potential approximation on the complex-time contour.
We have derived a set of new rules that supplement the celebrated Langreth
theorem and, as a whole, the generalized Langreth rules allow us to derive
NECPA equations for real time Green's functions. The solution of NECPA
equations provide the disorder averaged nonequilibrium density matrix as
well as other relevant quantities for quantum transport calculations. We
establish the excellent accuracy of NECPA by comparing its results to brute
force numerical calculations of disordered tight-binding models. Moreover,
the connection of NECPA equations which are derived on the complex-time
contour, to the nonequilibrium vertex correction theory which is derived on
the real-time axis, is made. As an application, we demonstrate that NECPA
can be combined with density functional theory to enable analysis of
nanoelectronic device physics from atomistic first principles.
\end{abstract}

\maketitle

\section{Introduction}

As dimensional scaling of electronic devices continues, quantum effects of
electron conduction is becoming increasingly important for practical design
of emerging systems. A theory typically starts from a given device
Hamiltonian from which quantum transport is analyzed by techniques such as
the scattering matrix and/or Keldysh nonequilibrium Green's function (NEGF)
methods\cite{Haug, Datta}. Calculation of device Hamiltonian under the
realistic condition of device operation is however a very complicated
problem and most device simulations rely on model and/or parameterized
Hamiltonians including the effective mass Hamiltonian, the $k\cdot p$ 
Hamiltonian, the tight-binding Hamiltonian, etc..
Combined with the NEGF formalism for quantum transport, these approaches
provide important insights for understanding nanoelectronic device physics.
However, there is a clear need in the physics community to advance atomistic
first principles, parameter-free and self-consistent methods to
fundamentally solve emerging nanoelectronic device problems. This is
necessary not only due to the lack of reliable phenomenological Hamiltonian
parameters for many materials and structures, but also due to the fact that
transport driven by an external bias is intrinsically a nonequilibrium
problem while parametrization of model Hamiltonian has usually been done at
equilibrium. Quantum transport theory at the atomistic level is also
necessary because the number of atoms in emerging generations of practical
devices is becoming countable.

One of the most basic requirements for any atomistic formalism of
nonequilibrium quantum transport is the ability to handle effects of
disorder. This is because all realistic device materials contain some degree
of unavoidable and random disorder such as atomic defects, vacancies,
surface roughness, interface irregularities, etc.. In addition, for many
situations the disorder is created by impurity doping in order to
functionalize the material such as semiconductors. Because the device
Hamiltonian depends on the configuration of the disorder, the predicted
physical properties must be averaged over the multitudes of disorder
configurations. Disorder average can be carried out by generating many
disorder configurations for a given disorder concentration. However, such
brute force analysis is computationally prohibitive in atomistic modeling of
realistic nano-devices. To overcome the prohibitively large computation
required for performing configuration average of disorder, it is desired to
obtain the averaged physical quantity without computing each impurity
configuration individually. In this regard, a well developed technique in
electronic structure theory is the coherent potential approximation (CPA)%
\cite{CPA1}. CPA is an effective medium technique by which the disorder
average of retarded Green's function $\overline{G^{r}}$ can be carried out
analytically. The CPA method is originally developed to study disordered bulk
materials\cite{CPA1} and later extended to investigate disorder effects in
surfaces and interfaces\cite{Turek}.

Recently significant progresses have been achieved to understand disorder
scattering in quantum transport by extending CPA with vertex correction
technique\cite{Velicky1}. In Ref.\onlinecite{Carva 2006}, Carva \textit{et al%
} calculated disorder averaged conductance in the linear response regime by
using vertex correction and the results were in good agreement to those of
supercell calculation. In Ref.\onlinecite{Ke 2008}, one of the authors and
collaborators advanced nonequilibrium vertex correction (NVC) theory which
applied vertex correction technique to NEGF and obtained disorder averaged
lesser Green's function $\overline{G^{<}}$ in addition to $\overline{G^{r}}$%
. Since important physical quantities in quantum transport can be expressed
in terms of $\overline{G^{r}}$ and $\overline{G^{<}}$ (see Section II), the
first principles CPA-NVC approach has been successfully applied to
investigate a variety of nonequilibrium quantum transport problems including
disorder effects in magnetic tunnel junctions\cite{Ke 2008,Ke 2010}, Cu
interconnects\cite{Ferdows 2010}, impurity limited mobility of short channel
graphene\cite{Zi 2012}, etc..

While the CPA-NVC theory is practically very useful, there are important
unresolved issues that require further theoretical investigation, for
instance, (i) Two different approximations, CPA for $\overline{G^{r}}$ and
NVC for $\overline{G^{<}}$, are used in the CPA-NVC theory. It is however
not proved that these two approximations are actually consistent with each
other at nonequilibrium, although it has been numerically verified as such
at equilibrium\cite{Ke 2008}. (ii) So far the NVC theory has been limited to
situations involving binary disorder sites, namely a site labeled $q$ can be
occupied by two species $A$ or $B$ with their respective statistical
weights. There are however many important nonequilibrium transport problems
that involve multiple species $q=A,B,C,\cdots $. How to extend CPA-NVC
theory to multiple species -- as can be done in CPA, is non-trivial and has
not been achieved. (iii) So far the NVC equation has been solved either
directly or iteratively. The direct solution requires solving an extremely
large linear equation array while the iterative solution is not always
numerically stable. It is necessary to develop a new method to solve $%
\overline{G^{<}}$ efficiently and smoothly, especially for systems with low
disorder concentration.

In this work, we shall develop a completely new approach\cite{RUC} other than the
vertex correction technique. The key idea is based on the fact that the
retarded Green's function and the contour ordered Green's function have the
same mathematical structure, as such the CPA equation for $\overline{G^{r}}$
can be viewed as the contour ordered CPA equation. By extending the
celebrated Langreth theorem\cite{Langreth}, we analytically continue from
the complex time contour to the real time axis such that the equations of
nonequilibrium coherent potential approximation (NECPA) are derived for both
$\overline{G^{r}}$ and $\overline{G^{<}}$ simultaneously. Although the NECPA
equations look very different from CPA-NVC equations, the two equation sets
can be proved to be equivalent to each other. The new formulation of NECPA
is not only elegant from the theoretical point of view, but also resolves
all the theoretical and practical issues mentioned in the last paragraph.
Finally, we present the combination of NECPA with the density functional
theory (DFT) to investigate nanoelectronic device physics from atomistic
first principles.

The rest of the paper is organized as follows. Section II gives a
brief review of NEGF formalism. Section III presents the theory of NECPA
which is the key work of this paper. Section IV discusses the connection of
NECPA to CPA-NVC. Section V provides numerical verifications of NECPA.
Section VI addresses the application of NECPA to DFT. Finally, the paper is
summarized in Section VII. Some mathematical proofs and technical details
are organized in several appendices.

\section{NEGF for quantum transport}

\label{sec-1}

To put the NECPA theory into context and for ease of presentations, in this
section we briefly review the NEGF formalism for calculating quantum
transport in two-probe systems with disorder. The formalism follows that of
Ref.\onlinecite{NEGF-jauho}.

\begin{figure}[tbp]
\includegraphics[width=1.4in, angle=90]{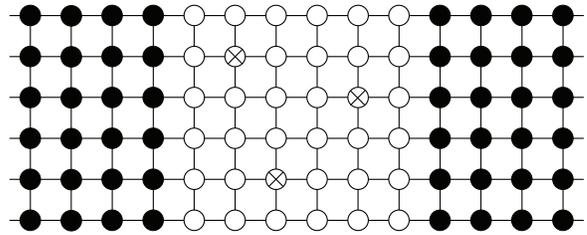}
\caption{Schematic plot of a two-probe system with some impurity sites in
the central scattering region. The left/right electrodes extend to $z=\pm
\infty $. The black dots are sites of electrodes, white circles are pure
sites in the scattering region, white sites with a cross are impurity sites.}
\label{fig1}
\end{figure}

Consider a general two-probe system consisting of a central scattering
region plus the left/right semi-infinite electrodes, schematically shown in
Fig.\ref{fig1}. There are disorder sites randomly located in the scattering
region indicated by the crossed-circles. Theoretically one may mimic
disorder effects by assigning the on-site energy to a random discrete
variable\cite{comment0}. It is assumed that on a disorder site-$i$, the
on-site energy $\varepsilon _{i}$ takes value $\varepsilon _{iq}$ with
probability $x_{iq}$, where $q=1,2,\cdots $ labels multiple impurity species
and, clearly, $\sum_{q}x_{iq}=1$.

The Hamiltonian of the two-probe system in the second quantization
representation can be written as:
\begin{eqnarray}
H &=&H_{C}+\sum_{\beta =L,R}H_{\beta }+\sum_{\beta =L,R}H_{C\beta }\ \ ,  \label{H1} \\
H_{C} &=&\sum_{i}\varepsilon _{i}c_{i}^{\dagger
}c_{i}+\sum_{i<j}t_{ij}c_{i}^{\dagger }c_{j}+t_{ij}^{\ast }c_{j}^{\dagger
}c_{i}\ \ ,  \label{HC1} \\
H_{\beta } &=&\sum_{k}\epsilon _{\beta k}a_{\beta k}^{\dagger }a_{\beta k}\
\ ,  \label{Hbeta} \\
H_{C\beta } &=&\sum_{ik}t_{ik}c_{i}^{\dagger }a_{\beta k}+t_{ik}^{\ast
}a_{\beta k}^{\dagger }c_{i}\ \ ,  \label{HCbeta}
\end{eqnarray}%
where $H_{C}$ is the Hamiltonian of the central scattering region, $H_{\beta
}$ ($\beta =L,R$) is the Hamiltonian of the left or right electrode, and $%
H_{C\beta }$ is the coupling between the scattering region and the $\beta $%
-electrode. Note that the above Hamiltonian is in a quadratic form thus
analytical solution of quantum transport can be obtained if the on-site
energy $\varepsilon _{i}$ is a definite variable. The complexity of our
problem comes from the fact that $\varepsilon _{i}$ is a random variable and
hence any physical quantities must be averaged over disorder configurations.

In the NEGF formalism, all physical quantities can be expressed in terms of
Green's functions. The most important quantities for a transport theory are
the electric current and the occupation number. The current flowing out of
the $\beta $-electrode can be derived as, (in atomic units $e=\hbar =1$)
\begin{equation}
I_{\beta }=2\text{Re}\int \frac{dE}{2\pi }\text{Tr }\left[ \overline{%
G^{r}\left( E\right) }\Sigma _{\beta }^{<}\left( E\right) +\overline{%
G^{<}\left( E\right) }\Sigma _{\beta }^{a}\left( E\right) \right] .
\label{current}
\end{equation}%
The occupation number of site-$i$ is calculated from the lesser Green's
function $G^{<}$,
\begin{equation}
N_{i}=\text{Im}\int \frac{dE}{2\pi }\text{Tr}\left[ \overline{G^{<}\left(
E\right) }\right] _{ii},  \label{Ni1}
\end{equation}%
in which $\left[ \cdots \right] _{ii}$ is to take the diagonal element
(diagonal block) of site-$i$. In these expressions, $G^{r}$ and $G^{<}$ are
the retarded and lesser Green's functions of the central region of the
system, and $\Sigma _{\beta }^{<}$ and $\Sigma _{\beta }^{a}$ are the lesser
and advanced self-energies of the $\beta $-electrode. The notation $%
\overline{(\cdots )}$ means these quantities need to be averaged over
disorder configurations $\left\{ \varepsilon _{i}\right\} $. The advanced
Green's function and the advanced self-energy are Hermitian conjugates of
their retarded counterparts,
\begin{eqnarray*}
G^{a}\left( E\right)  &=&\left[ G^{r}\left( E\right) \right] ^{\dagger }, \\
\Sigma _{\beta }^{a}\left( E\right)  &=&\left[ \Sigma _{\beta }^{r}\left(
E\right) \right] ^{\dagger }.
\end{eqnarray*}

To proceed further, $G^{r}$ and $G^{<}$ are solved by Dyson equation and
Keldysh equation, respectively,
\begin{equation}
G^{r}\left( E\right) =\left[ E-H_{C}^{0}-\varepsilon -\Sigma ^{r}\left(
E\right) \right] ^{-1},  \label{GR1}
\end{equation}%
\begin{equation}
G^{<}\left( E\right) =G^{r}\left( E\right) \Sigma ^{<}\left( E\right)
G^{a}\left( E\right) ,  \label{Gless1}
\end{equation}%
where $H_{C}^{0}$ is the off-diagonal (definite) part of $H_{C}$, $%
\varepsilon \equiv diag\left( \left[ \varepsilon _{1},\varepsilon
_{2},\cdots \right] \right) $ is the diagonal (random) part of $H_{C}$, and $%
\Sigma ^{r}\left( E\right) $ and $\Sigma ^{<}\left( E\right) $ are the
retarded and lesser total self-energies
\begin{equation}
\Sigma ^{r}\left( E\right) =\Sigma _{L}^{r}\left( E\right) +\Sigma
_{R}^{r}\left( E\right) ,  \label{Sigma1}
\end{equation}%
\begin{equation}
\Sigma ^{<}\left( E\right) =\Sigma _{L}^{<}\left( E\right) +\Sigma
_{R}^{<}\left( E\right) ,  \label{Sigma-less}
\end{equation}%
and
\begin{equation}
\Sigma _{\beta }^{<}\left( E\right) =f_{\beta }\left( E\right) \left[ \Sigma
_{\beta }^{a}\left( E\right) -\Sigma _{\beta }^{r}\left( E\right) \right] .
\label{Sbeta1}
\end{equation}%
Here $f_{\beta }\left( E\right) $ is the Fermi function of the $\beta $%
-electrode. Note we have assumed that all disorder sites are located in the
central region and electrodes have no disorder. Otherwise one can always
enlarge the central region to include all disorder sites\cite{comment1}.
With this in mind, the disorder average is done to the Green's functions and
not to the self-energies of the electrodes.

Another important physical quantity in quantum transport is the transmission
coefficient $T$. Note that the electric current in Eq.(\ref{current}) can be
rewritten as
\begin{equation}
I_{L}=-I_{R}=\int \frac{dE}{2\pi }T\left( E\right) \left[ f_{L}\left(
E\right) -f_{R}\left( E\right) \right] ,  \notag
\end{equation}%
in which $T(E)$ is the transmission coefficient which can be expressed in
terms of Green's functions,
\begin{equation}
T\left( E\right) =\text{Tr }\overline{G^{r}\left( E\right) \Gamma _{L}\left(
E\right) G^{a}\left( E\right) \Gamma _{R}\left( E\right) },  \label{T1}
\end{equation}%
in which $\Gamma _{\beta }\left( E\right) \equiv -i\left[ \Sigma _{\beta
}^{a}\left( E\right) -\Sigma _{\beta }^{r}\left( E\right) \right] $ is the
line-width function of the $\beta $-electrode. Notice that $\Sigma
^{<}\left( E\right) $ is reduced to $\Gamma _{L}\left( E\right) $ by making
the substitution $f_{L}\left( E\right) \rightarrow -i$ and $f_{R}\left(
E\right) \rightarrow 0$ in Eq.(\ref{Sigma-less}) and Eq.(\ref{Sbeta1}). So
the calculation of disorder averaged transmission $T\left( E\right) $ can be
reduced to the calculation of $\overline{G^{<}}$,
\begin{eqnarray}
T\left( E\right)  &=&\text{Tr }\overline{G^{r}\left( E\right) \Gamma
_{L}\left( E\right) G^{a}\left( E\right) }\Gamma _{R}\left( E\right)   \notag
\\
&=&\text{Tr }\overline{G_{L}^{<}\left( E\right) }\Gamma _{R}\left( E\right) ,
\label{T2}
\end{eqnarray}%
in which $\overline{G_{L}^{<}\left( E\right) }$ is defined as%
\begin{equation*}
\overline{G_{L}^{<}\left( E\right) }\equiv \left[ \overline{G^{<}\left(
E\right) }\right] _{f_{L}\left( E\right) \rightarrow -i,f_{R}\left( E\right)
\rightarrow 0}.
\end{equation*}

With these expressions, all the analysis of disorder effects are reduced to
evaluate $\overline{G^{r}}$ and $\overline{G^{<}}$ (hereafter the argument $E
$ is omitted for simplicity of notations). In principle, by brute force one
can calculate the Green's functions for each disorder configuration and
evaluate the average afterward. As mentioned in the Introduction, such a
brute force calculation quickly becomes formidable due to the huge number of
configurations that scales up exponentially with the number of disorder
sites in the two-probe systems. We therefore seek for the approximation
technique to evaluate the average analytically.

\section{Theory of NECPA}

Having discussed the general formalism of NEGF for quantum transport, in
this section we present NECPA theory for disordered two-probe systems.

\begin{figure}[tbp]
\includegraphics[height=8cm, angle=90]{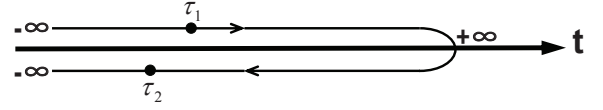}
\caption{Schematic plot of the complex-time contour that goes above the real
time axis from $\protect\tau =-\infty $ to $\protect\tau =+\infty $, and
returns below the real time axis to $\protect\tau =-\infty $. The contour
ordered Green's function is defined on the complex-time contour. The NECPA
equations are derived by analytic continuation of the contour ordered CPA
equation. }
\label{fig2}
\end{figure}

In many-body theory\cite{Baym}, it is well known that nonequilibrium
statistics can be formulated via the contour ordered Green's function $%
G(\tau _{1},\tau _{2})$ where $\tau _{1}$ and $\tau _{2}$ are the
complex-time, as shown in Fig.\ref{fig2}. The key idea of NECPA is to carry
out disorder average within CPA to contour ordered Green's function to
obtain $\overline{G(\tau _{1},\tau _{2})}$. Afterward we transform the
disorder averaged contour ordered Green's function $\overline{G}$ to real
time Green's functions $\overline{G^{r}}$ and $\overline{G^{<}}$ by analytic
continuation.

In the following subsections, we first introduce contour ordered CPA
equation which is the starting point of NECPA theory. We proceed to derive
NECPA equations by applying generalized Langreth theorem to contour ordered
CPA equation. After that we discuss the iterative method for solving NECPA
equations and the analytical solutions in the low disorder concentration
limit. Finally we investigate a special but important case, namely two-probe
systems with transverse periodicity.

\subsection{Contour ordered CPA equation}

The spirit of CPA is to replace the random on-site energy $\{\varepsilon
_{i}\}$ in Eq.(\ref{GR1}) by an effective on-site energy $\{\tilde{%
\varepsilon}_{i}^{r}\}$ such that the average scattering vanishes with
respect to the effective media. Originally CPA was developed for the
retarded Green's function (see e.g., Eq.(9) of the first paper in Ref.%
\onlinecite{CPA1}):
\begin{equation}
\left\{
\begin{array}{c}
\overline{t_{i}^{r}}=\sum_{q}x_{iq}t_{iq}^{r}=0, \\
\\
t_{iq}^{r}\equiv \left[ \left( \varepsilon _{iq}-\tilde{\varepsilon}%
_{i}^{r}\right) ^{-1}-\overline{G_{i}^{r}}\right] ^{-1}, \\
\\
\overline{G_{i}^{r}}\equiv \left[ \overline{G^{r}}\right] _{ii}, \\
\\
\overline{G^{r}}=\left[ E-H_{C}^{0}-\tilde{\varepsilon}^{r}-\Sigma ^{r}%
\right] ^{-1},%
\end{array}%
\right.   \label{CPA_r}
\end{equation}%
where $\tilde{\varepsilon}^{r}\equiv diag\left( \left[ \tilde{\varepsilon}%
_{1}^{r},\tilde{\varepsilon}_{2}^{r},\cdots \right] \right) $ is a diagonal
matrix of the effective on-site energies which is called the coherent
potential in the literature\cite{CPA1}. The first line in Eq.(\ref{CPA_r})
states the sprit of CPA, namely on the disorder site-$i$ scattering
processes due to various impurity species cancel with each other.

Now we extend the idea of CPA to the contour ordered Green's function. Since
the contour ordered Green's function and retarded Green's function satisfy
the same equation of motion and hence have the same mathematical structure%
\cite{Baym,Haug}, it is straightforward to write the contour ordered
CPA equation by simply removing the superscript $r$ in Eq.(\ref{CPA_r}), we
have,
\begin{equation}
\left\{
\begin{array}{c}
\overline{t_{i}}=\sum_{q}x_{iq}t_{iq}=0, \\
\\
t_{iq}\equiv \left[ \left( \varepsilon _{iq}-\tilde{\varepsilon}_{i}\right)
^{-1}-\overline{G_{i}}\right] ^{-1}, \\
\\
\overline{G_{i}}\equiv \left[ \overline{G}\right] _{ii}, \\
\\
\overline{G}=\left[ E-H_{C}^{0}-\tilde{\varepsilon}-\Sigma \right] ^{-1},%
\end{array}%
\right.  \label{CPA_tao}
\end{equation}%
where $\overline{t_{i}}$, $t_{iq}$, $\tilde{\varepsilon}_{i}$, $\overline{%
G_{i}}$, $\overline{G}$, $\Sigma $ are defined on the complex-time contour
and have been Fourier transformed to frequency space.

It is more intuitive to rewrite Eq.(\ref{CPA_tao}) into an equivalent form
by using the conditional Green's function:
\begin{equation}
\left\{
\begin{array}{c}
\overline{G_{i}}=\sum_{q}x_{iq}\overline{G_{iq}}, \\
\\
\overline{G_{i}}=\left[ \left( E-H_{C}^{0}-\tilde{\varepsilon}-\Sigma
\right) ^{-1}\right] _{ii}, \\
\\
\overline{G_{iq}}=\left[ \left( E-H_{C}^{0}-\tilde{\varepsilon}^{iq}-\Sigma
\right) ^{-1}\right] _{ii}%
\end{array}%
\right.  \label{CPA_conditional}
\end{equation}%
where $\tilde{\varepsilon}^{iq}$ means to replace the $i$-th diagonal
element of $\tilde{\varepsilon}$ by $\varepsilon _{iq}$. $\overline{G_{iq}}$
is the conditional Green's function, namely the Green's function of site-$i$
under the condition that the site is occupied by specie-$q$ and other sites
remain disorder sites. The first line in Eq.(\ref{CPA_conditional}) is
consistent with the meaning of conditional Green's function. The derivation
of the first line in Eq.(\ref{CPA_conditional}) can be found in the Appendix-%
\ref{maths}.

Finally, Eq.(\ref{CPA_conditional}) can be further reduced to another
convenient but equivalence form:%
\begin{equation}
\left\{
\begin{array}{c}
\overline{G_{i}}=\sum_{q}x_{iq}\overline{G_{iq}}, \\
\\
\overline{G}=\left[ E-H_{C}^{0}-\tilde{\varepsilon}-\Sigma \right] ^{-1}, \\
\\
\overline{G_{i}}=\left[ \overline{G}\right] _{ii}, \\
\\
\overline{G_{i}}=\left[ E-\tilde{\varepsilon}_{i}-\Omega _{i}\right] ^{-1},
\\
\\
\overline{G_{iq}}=\left[ E-\varepsilon _{iq}-\Omega _{i}\right] ^{-1},%
\end{array}%
\right.  \label{CPA_Omega}
\end{equation}%
where $\Omega _{i}$ is the contour ordered coherent interactor\cite%
{Kudrnovsky}. Eq.(\ref{CPA_Omega}) will be used in the analytic continuation
in the next subsection. The derivation of the fourth and fifth lines in Eq.(%
\ref{CPA_Omega}) can be found in the Appendix-\ref{maths}.

\subsection{Generalized Langreth theorem and NECPA equations}

In this subsection, we shall apply analytic continuation to the contour ordered
CPA equation (\ref{CPA_Omega}) to obtain disorder averaged real time Green's
function. In this regard, one usually applies the Langreth theorem\cite%
{Langreth,Haug} that bridges between the contour ordered Green's function $G$
and the real-time Green's functions $G^{r,a}$ and $G^{<,>}$. According to
the Langreth theorem, if contour ordered quantities $A$, $B$, $C$ satisfy $%
C=AB$, then the corresponding retarded and lesser quantities are obtained as:%
\begin{equation}
C^{r}=A^{r}B^{r},  \label{Langreth r}
\end{equation}

\begin{equation}
C^{<}=A^{r}B^{<}+A^{<}B^{a}.  \label{Langreth d}
\end{equation}%
In addition, if a contour ordered quantity $D$ does not have a finite
imaginary part so that $D^{r}$ and $D^{a}$ are indistinguishable, it behaves
as a constant,
\begin{eqnarray}
D^{r} &=&D^{a}=D,  \label{Langreth r3} \\
D^{<} &=&0.  \label{Langreth d3}
\end{eqnarray}%
Looking at the form of the contour ordered CPA equation (\ref{CPA_Omega}),
it is obvious that the above Langreth rules Eqs.(\ref{Langreth r},\ref%
{Langreth d},\ref{Langreth r3},\ref{Langreth d3}) cannot be directly applied
because the right hand side of the second, fourth and fifth lines in Eq.(\ref%
{CPA_Omega}) involve the inverse operations. Therefore a new set of Langreth
rules need to established to determine quantities like $(A^{-1})^{r,a,<,>}$.
This can be accomplished as follows. Let $C=AA^{-1}=1$ and apply
Eqs.(\ref{Langreth r},\ref{Langreth d},\ref{Langreth r3},\ref{Langreth d3}),
two new rules for the inverse are derived as follows:
\begin{eqnarray}
\left( A^{-1}\right) ^{r} &=&\left( A^{r}\right) ^{-1},  \label{Langreth r2}
\\
\left( A^{-1}\right) ^{<} &=&-\left( A^{r}\right) ^{-1}A^{<}\left(
A^{a}\right) ^{-1}.  \label{Langreth d2}
\end{eqnarray}%
These rules allow one to carry out analytic continuation of the inverse of
contour ordered quantities. The two new Langreth rules Eqs.(\ref{Langreth r2}%
,\ref{Langreth d2}), together with the original rules Eqs.(\ref{Langreth r},%
\ref{Langreth d},\ref{Langreth r3},\ref{Langreth d3}), shall be referred to
as generalized Langreth theorem in the rest of this paper.

By applying the generalized Langreth theorem to the contour ordered CPA
equation (\ref{CPA_Omega}), two sets of equations can be obtained for $%
\overline{G^{r}}$ and $\overline{G^{<}}$ respectively:

\begin{equation}
\left\{
\begin{array}{c}
\overline{G_{i}^{r}}=\sum_{q}x_{iq}\overline{G_{iq}^{r}}, \\
\\
\overline{G^{r}}=\left[ E-H_{C}^{0}-\tilde{\varepsilon}^{r}-\Sigma ^{r}%
\right] ^{-1}, \\
\\
\overline{G_{i}^{r}}=\left[ \overline{G^{r}}\right] _{ii}, \\
\\
\overline{G_{i}^{r}}=\left[ E-\tilde{\varepsilon}_{i}^{r}-\Omega _{i}^{r}%
\right] ^{-1}, \\
\\
\overline{G_{iq}^{r}}=\left[ E-\varepsilon _{iq}-\Omega _{i}^{r}\right]
^{-1},%
\end{array}%
\right.  \label{NECPA Gr}
\end{equation}
\begin{equation}
\left\{
\begin{array}{c}
\overline{G_{i}^{<}}=\sum_{q}x_{iq}\overline{G_{iq}^{<}}, \\
\\
\overline{G^{<}}=\overline{G^{r}}\left( \Sigma ^{<}+\tilde{\varepsilon}%
^{<}\right) \overline{G^{a}}, \\
\\
\overline{G_{i}^{<}}=\left[ \overline{G^{<}}\right] _{ii}, \\
\\
\overline{G_{i}^{<}}=\overline{G_{i}^{r}}\left( \tilde{\varepsilon}%
_{i}^{<}+\Omega _{i}^{<}\right) \overline{G_{i}^{a}}, \\
\\
\overline{G_{iq}^{<}}=\overline{G_{iq}^{r}}\Omega _{i}^{<}\overline{%
G_{iq}^{a}}.%
\end{array}%
\right.  \label{NECPA Gd}
\end{equation}

These two equations are the central results of this work which extends the
equilibrium CPA for bulk systems to the nonequilibrium two-probe systems. As
expected the equation of $\overline{G^{r}}$ recovers the known CPA equation,
and $\overline{G^{<}}$ will be shown to be equivalent to the NVC equation in Ref.%
\onlinecite{Ke 2008}. This way, by applying the generalized Langreth theorem
to the contour ordered CPA equation, both $\overline{G^{r}}$ and $\overline{%
G^{<}}$ are derived simultaneously. Collectively, in the rest of this paper
we shall refer to Eq.(\ref{NECPA Gr}) and Eq.(\ref{NECPA Gd}) as the NECPA
equations.

The NECPA theory presented above has several distinct advances at both the
fundamental level and the practical level: (i) NECPA treats disorder average
for $\overline{G^{r}}$ and $\overline{G^{<}}$ on equal footing and the
derived equations are similar in form. (ii) NECPA derives the averaged
conditional Green's functions $\overline{G_{iq}^{r}}$ and $\overline{%
G_{iq}^{<}}$ for disorder site with any number of impurity species. (iii)
NECPA provides a natural iterative method for solving $\overline{G^{r}}$ and
$\overline{G^{<}}$ which will be the subject of the next subsection.

\subsection{Solving the NECPA equations}

NECPA equations not only make a theoretical advance but also provide a
natural iterative method for solving $\overline{G^{r}}$ and $\overline{G^{<}}
$. With the aid of Eq.(\ref{NECPA Gr}), $\overline{G^{r}}$ can be solved
with following iterative method:

\begin{enumerate}
\item Make an initial guess of $\Omega ^{r}$.

\item Determine $\tilde{\varepsilon}^{r}$ from the first, the fourth and the
fifth lines of Eq.(\ref{NECPA Gr}), the result is:
\begin{equation*}
\tilde{\varepsilon}_{i}^{r}=E-\Omega _{i}^{r}-\left[ \sum_{q}x_{iq}\left(
E-\varepsilon _{iq}-\Omega _{i}^{r}\right) ^{-1}\right] ^{-1}.
\end{equation*}

\item Determine $\overline{G^{r}}$ from the second line of Eq.(\ref{NECPA Gr}%
)%
\begin{equation*}
\overline{G^{r}}=\left[ E-H_{C}^{0}-\tilde{\varepsilon}^{r}-\Sigma ^{r}%
\right] ^{-1}.
\end{equation*}

\item Update $\Omega ^{r}$ by solving it from the fourth line of Eq.(\ref%
{NECPA Gr}), the result is:
\begin{equation*}
\Omega _{i}^{r}=E-\tilde{\varepsilon}_{i}^{r}-\left[ \overline{G^{r}}\right]
_{ii}^{-1}\ \ .
\end{equation*}

\item Go back to step-2 to repeat the process until $\Omega ^{r}$ is fully
converged.
\end{enumerate}

With the aid of Eq.(\ref{NECPA Gd}), $\overline{G^{<}}$ can be solved with
following iterative method:

\begin{enumerate}
\item Make an initial guess of $\Omega ^{<}$.

\item Determine $\tilde{\varepsilon}^{<}$ from the first, the fourth and the
fifth lines of Eq.(\ref{NECPA Gd}), the result is:
\begin{equation*}
\tilde{\varepsilon}_{i}^{<}=\left( \overline{G_{i}^{r}}\right) ^{-1}\left[
\sum_{q}x_{iq}\overline{G_{iq}^{r}}\Omega _{i}^{<}\overline{G_{iq}^{a}}%
\right] \left( \overline{G_{i}^{a}}\right) ^{-1}-\Omega _{i}^{<}\ \ .
\end{equation*}

\item Determine $\overline{G^{<}}$ from the second line of Eq.(\ref{NECPA Gd}%
)%
\begin{equation*}
\overline{G^{<}}=\overline{G^{r}}\left( \Sigma ^{<}+\tilde{\varepsilon}%
^{<}\right) \overline{G^{a}}.
\end{equation*}

\item Update $\Omega ^{<}$ by solving it from the fourth line of Eq.(\ref%
{NECPA Gd}), the result is:
\begin{equation}
\Omega _{i}^{<}=\left( \overline{G_{i}^{r}}\right) ^{-1}\left[ \overline{%
G^{<}}\right] _{ii}\left( \overline{G_{i}^{a}}\right) ^{-1}-\tilde{%
\varepsilon}_{i}^{<}\ \ .  \notag
\end{equation}

\item Go back to step-2 to repeat the process until $\Omega ^{<}$ is fully
converged.
\end{enumerate}

Note that quantities $\overline{G^{r}}$, $\overline{G^{a}}$, $\overline{%
G_{iq}^{r}}$ and $\overline{G_{iq}^{a}}$ are assumed to be known in the
iterative solution to $\overline{G^{<}}$. In practice, the iterations of $%
\overline{G^{r}}$ and $\overline{G^{<}}$ are actually carried out together.

It is interesting to analyze the computational cost of the above methods.
Assume that there are $N_{D}$ disorder sites in the central region of a
two-probe system (see Fig.\ref{fig1}). The costs of step-2 and step-4 are
proportional to $N_{D}$, while the cost of step-3 is proportional to $%
N_{D}^{3}$ if full matrix operations are used. So the \emph{bottleneck} of
the iteration is step-3 which needs to be carefully optimized. Notice that
in two-probe systems the size along the transport dimension is usually much
larger than the transverse dimensions. Taking advantage of this geometry,
the cost of step-3 can be drastically reduced using the principal layer
approach discussed in the Appendix-\ref{PL-subsec}.

The computational cost for solving the NECPA equations can be further
reduced if the disorder concentration is very low. In typical semiconductor
devices a doping concentration of $10^{20}cm^{-3}$ (heavily doped) amounts
to a disorder concentration $x\sim 2\times 10^{-3}$. For such low disorder
concentration $x$, the solution to NECPA equations can be approximated to
high precision by analytical expressions obtained by perturbation expansion
with respect to the small parameter $x$. Let $q=0$ label the host material
specie and $q>0$ label impurity species. Low disorder concentration means $%
x_{i,q=0}\gg x_{i,q>0}$. We also have $\sum_{q}x_{iq}=1$ due to
normalization. The solution of NECPA equations can be obtained up to the
first order of $x_{i,q>0}$:
\begin{equation}
\tilde{\varepsilon}_{i}^{r}\thickapprox \varepsilon
_{i0}+\sum_{q>0}x_{iq}t_{iq}^{r},  \label{lowX Gr}
\end{equation}%
\begin{equation}
\tilde{\varepsilon}_{i}^{<}\thickapprox
\sum_{q>0}x_{iq}t_{iq}^{r}G_{0,ii}^{<}t_{iq}^{a},  \label{lowX Gd}
\end{equation}%
in which
\begin{equation}
t_{iq}^{r}=\left[ \left( \varepsilon _{iq}-\varepsilon _{i0}\right)
^{-1}-G_{0,ii}^{r}\right] ^{-1},
\end{equation}%
\begin{equation}
G_{0}^{r}=\left[ E-H_{C}^{0}-\varepsilon ^{0}-\Sigma ^{r}\right] ^{-1},
\end{equation}%
\begin{equation}
G_{0}^{<}=G_{0}^{r}\Sigma ^{<}G_{0}^{a},
\end{equation}%
where $\varepsilon ^{0}=diag\left( \left[ \varepsilon _{10},\varepsilon
_{20},\cdots \right] \right) $ is the on-site energy of host material. These
analytical expressions allow one to calculate $\overline{G^{r}}$ and $%
\overline{G^{<}}$ without the iteration procedure discussed above. Using the
first order (in $x$) formula, we found that the total computational cost of transport
is roughly twice that of the corresponding clean system.

\subsection{NECPA with transverse periodicity}

\label{periodic-subsec}

In some applications one can identify small unit cell in the transverse
dimensions of two-probe systems. For two-probe systems without disorder, the
transverse periodicity allows one to apply the Bloch theorem and make $k$%
-sampling in the Brillouin zone. Thus the calculation of the transverse periodic
two-probe system is reduced to the calculation in a small unit cell plus $k$%
-sampling. For two-probe systems with random disorder, the translational
symmetry is broken in the transverse dimensions and Bloch theorem does not
hold. Nevertheless, NECPA is an effective medium theory whose application
restores the translational symmetry of $\overline{G^{r}}$ and $\overline{%
G^{<}}$. Therefore one can still work with the small unit cell plus $k$%
-sampling to calculate $\overline{G^{r}}$ and $\overline{G^{<}}$.

For transverse periodic two-probe systems, NECPA equations need to be
modified slightly to include $k$-sampling:
\begin{equation}
\left\{
\begin{array}{c}
\overline{G_{i}^{r}}=\sum_{q}x_{iq}\overline{G_{iq}^{r}}, \\
\\
\overline{G^{r}}\left( k\right) =\left[ E-H_{C}^{0}\left( k\right) -\tilde{%
\varepsilon}^{r}-\Sigma ^{r}\left( k\right) \right] ^{-1}, \\
\\
\overline{G^{r}}=\int_{-\pi }^{+\pi }\frac{dk}{2\pi }\overline{G^{r}}\left(
k\right) , \\
\\
\overline{G_{i}^{r}}=\left[ \overline{G^{r}}\right] _{ii}, \\
\\
\overline{G_{i}^{r}}=\left[ E-\tilde{\varepsilon}_{i}^{r}-\Omega _{i}^{r}%
\right] ^{-1}, \\
\\
\overline{G_{iq}^{r}}=\left[ E-\varepsilon _{iq}-\Omega _{i}^{r}\right]
^{-1}.%
\end{array}%
\right.  \label{NECPA Gr periodic}
\end{equation}%
\begin{equation}
\left\{
\begin{array}{c}
\overline{G_{i}^{<}}=\sum_{q}x_{iq}\overline{G_{iq}^{<}}, \\
\\
\overline{G^{<}}\left( k\right) =\overline{G^{r}}\left( k\right) \left[
\Sigma ^{<}\left( k\right) +\tilde{\varepsilon}^{<}\right] \overline{G^{a}}%
\left( k\right) , \\
\\
\overline{G^{<}}=\int_{-\pi }^{+\pi }\frac{dk}{2\pi }\overline{G^{<}}\left(
k\right) , \\
\\
\overline{G_{i}^{<}}=\left[ \overline{G^{<}}\right] _{ii}, \\
\\
\overline{G_{i}^{<}}=\overline{G_{i}^{r}}\left( \tilde{\varepsilon}%
_{i}^{<}+\Omega _{i}^{<}\right) \overline{G_{i}^{a}}, \\
\\
\overline{G_{iq}^{<}}=\overline{G_{iq}^{r}}\Omega _{i}^{<}\overline{%
G_{iq}^{a}}.%
\end{array}%
\right.  \label{NECPA Gd periodic}
\end{equation}%
In the above equations, $H_{C}^{0}\left( k\right) $ and $\Sigma ^{r,<}\left(
k\right) $ are the Fourier transform of $H_{C}^{0}$ and $\Sigma ^{r,<}$. $k$
is the dimensionless wave vector: For systems with periodicity in one
transverse dimension, $k$ is defined as $\mathbf{k\cdot a}$ in which $%
\mathbf{k}$ is the wave vector and $\mathbf{a}$ is the unit cell vector of
the periodic transverse dimension. For systems with periodicity in two
transverse dimensions,$\ k$ is defined as $\left( k_{1},k_{2}\right) =(%
\mathbf{k\cdot a}_{1},\mathbf{k\cdot a}_{2})$ in which $\mathbf{a}_{1}$ and $%
\mathbf{a}_{2}$ are the two unit cell vectors of the periodic transverse
dimensions. Correspondingly, $\int_{-\pi }^{+\pi }\frac{dk}{2\pi }$should be
understood as $\int_{-\pi }^{+\pi }\frac{dk_{1}}{2\pi }\int_{-\pi }^{+\pi }%
\frac{dk_{2}}{2\pi }$.

\section{Further discussions}

This section is devoted to establish the connection of the newly developed
NECPA theory to the existing CPA-NVC theory\cite{Ke 2008}. Three issues are
discussed in the following subsections: the equivalence of NECPA theory and
CPA-NVC theory, a disorder averaged Ward-type identity, and conditional
Green's function in binary systems.

\subsection{Equivalence of NECPA and CPA-NVC}

In CPA-NVC theory, $\overline{G^{<}}$ is calculated by using the NVC technique
where $\overline{G^{<}}$ is decomposed into two parts, a simple averaged
term and a vertex correlation term, namely
\begin{eqnarray}
\overline{G^{<}} &=&\overline{G^{r}\Sigma ^{<}G^{a}},  \notag \\
&=&\overline{G^{r}}\Sigma ^{<}\overline{G^{a}}+\overline{G^{r}}\Lambda
\overline{G^{a}},  \label{NVC 1}
\end{eqnarray}%
in which $\Lambda $ is a diagonal matrix $diag\left( \left[ \Lambda
_{1},\Lambda _{2},\cdots \right] \right) $ and is referred to as
nonequilibrium vertex correction. $\Lambda $ satisfies the following NVC
equation\cite{Ke 2008},
\begin{eqnarray}
\Lambda _{i} &=&\sum_{q}x_{iq}t_{iq}^{r}\left[ \overline{G^{r}}\left( \Sigma
^{<}+\Lambda \right) \overline{G^{a}}\right] _{ii}t_{iq}^{a}-  \notag \\
&&\sum_{q}x_{iq}t_{iq}^{r}\overline{G_{i}^{r}}\Lambda _{i}\overline{G_{i}^{a}%
}t_{iq}^{a}\ \ ,  \label{NVC 2}
\end{eqnarray}%
in which $t_{iq}^{r}\equiv \left[ \left( \varepsilon _{iq}-\tilde{\varepsilon%
}_{i}^{r}\right) ^{-1}-\overline{G_{i}^{r}}\right] ^{-1}$. It is not obvious
at all that $\overline{G^{<}}$ solved from Eqs.(\ref{NVC 1},\ref{NVC 2}) are
the same as the solution of Eq.(\ref{NECPA Gd}) in the NECPA theory.

The differences between NECPA theory and CPA-NVC theory come from distinct
theoretical paths. In the CPA-NVC theory, one starts from contour ordered $G$%
, and derive the Dyson equation (\ref{GR1}) for $G^{r}$ and the Keldysh
equation (\ref{Gless1}) for $G^{<}$ by analytic continuation. Afterward
disorder average is carried out with CPA and NVC techniques to obtain $%
\overline{G^{r}}$ and $\overline{G^{<}}$. In the NECPA theory, one also
starts from the contour ordered $G$, but the disorder average is carried out
by CPA to obtain $\overline{G}$ before the analytic continuation. Afterward $%
\overline{G^{r}}$ and $\overline{G^{<}}$ are derived on equal footing by
applying analytic continuation to $\overline{G}$. Therefore $\overline{G^{<}}
$ derived by the two theories must be equivalent although the mathematical
forms look very different. Explicitly, we are able to prove the equivalence
by showing that the nonequilibrium vertex correction $\Lambda $ in CPA-NVC
theory is actually identical to the lesser coherent potential $\tilde{\varepsilon%
}^{<}$ in NECPA theory. The proof is presented in Appendix-\ref{app-A}.

\subsection{Disorder averaged Ward-type identity}

In the NEGF formalism, there is a Ward-type identity which links $G^{r,a}$
to $G^{<,>}$,
\begin{equation}
G^{r}-G^{a}\equiv G^{>}-G^{<}.  \label{Ward1}
\end{equation}%
By disorder averaging on both sides, one obtains:
\begin{equation}
\overline{G^{r}}-\overline{G^{a}}\equiv \overline{G^{>}}-\overline{G^{<}}.
\label{CPA-NVC}
\end{equation}%
If the disorder average is done \emph{rigorously}, this disorder averaged
identity is obviously true. However, in CPA-NVC theory, approximation
techniques are used to carry out disorder average: CPA is applied to the
left hand side of Eq.(\ref{CPA-NVC}) while NVC to the right hand side. It
was shown by numerical computation at equilibrium\cite{Ke 2008} that Eq.(\ref%
{CPA-NVC}) holds true to extremely high precision for many disordered
structures. It is indeed amazing that the equality holds even after making
approximations on the two sides\cite{comment3}.

From the point of view of NECPA, the identity Eq.(\ref{CPA-NVC}) is however
self-evident: coherent potential approximation is made to the contour
ordered Green's function and afterward all real-time Green's functions $%
\overline{G^{r,a}}$ and $\overline{G^{<,>}}$ are derived from analytic
continuation without further approximation. That is why the identity still
holds even after disorder average. In Appendix-\ref{app-A}, we have proved
the equivalence of NECPA theory and CPA-NVC theory which indirectly proves
that CPA and NVC are consistent approximations. Finally, we would like to
mention that the disorder averaged identity is very useful to test codes in
the numerical implementation.

\subsection{Conditional Green's functions for binary systems}

\label{binary-subsec}

For systems having binary disorder sites, i.e., $q=A,B$, there is an
alternative method -- the random variable method, to partition $\overline{%
G_{i}^{r}}$ and $\overline{G_{i}^{<}}$ into conditional Green's function $%
\overline{G_{iq}^{r}}$ and $\overline{G_{iq}^{<}}$. Define random variables $%
\eta _{iA}$ and $\eta _{iB}$: $\eta _{iq}=1$ if site-$i$ is occupied by
specie-$q$ and $\eta _{iq}=0$ otherwise. It follows that $\overline{\eta
_{iA}}=x_{iA}$, $\overline{\eta _{iB}}=x_{iB}$, and
\begin{eqnarray*}
\eta _{iA}+\eta _{iB} &=&1, \\
\eta _{iA}\varepsilon _{iA}+\eta _{iB}\varepsilon _{iB} &=&\varepsilon _{i}.
\end{eqnarray*}%
By definition, $\overline{G_{iq}^{r}}$ and $\overline{G_{iq}^{<}}$ can be
expressed in terms of $\eta _{iq}$%
\begin{eqnarray*}
\overline{G_{iq}^{r}} &=&\frac{\overline{\eta _{iq}G_{i}^{r}}}{\overline{%
\eta _{iq}}}, \\
\ \ \ \ \ \overline{G_{iq}^{<}} &=&\frac{\overline{\eta _{iq}G_{i}^{<}}}{%
\overline{\eta _{iq}}},
\end{eqnarray*}%
where $G_{i}^{r}=\left[ G^{r}\right] _{ii}$ and $G_{i}^{<}=\left[ G^{<}%
\right] _{ii}$.

After some algebra\cite{comment2}, $\overline{G_{iq}^{r}}$ and $\overline{%
G_{iq}^{<}}$ can be derived as
\begin{eqnarray}
\overline{G_{iA}^{r}} &=&\frac{1}{x_{iA}}\left( \varepsilon
_{iA}-\varepsilon _{iB}\right) ^{-1}\left( \tilde{\varepsilon}%
_{i}^{r}-\varepsilon _{iB}\right) \overline{G_{i}^{r}}\ ,  \notag \\
\overline{G_{iB}^{r}} &=&\frac{1}{x_{iB}}\left( \varepsilon
_{iB}-\varepsilon _{iA}\right) ^{-1}\left( \tilde{\varepsilon}%
_{i}^{r}-\varepsilon _{iA}\right) \overline{G_{i}^{r}}\ ,  \label{Grq}
\end{eqnarray}%
\begin{eqnarray}
\overline{G_{iA}^{<}} &=&\frac{1}{x_{iA}}\left( \varepsilon
_{iA}-\varepsilon _{iB}\right) ^{-1}\left[ \left( \tilde{\varepsilon}%
_{i}^{r}-\varepsilon _{iB}\right) \overline{G_{i}^{<}}+\tilde{\varepsilon}%
_{i}^{<}\overline{G_{i}^{a}}\right] ,  \notag \\
\overline{G_{iB}^{<}} &=&\frac{1}{x_{iB}}\left( \varepsilon
_{iB}-\varepsilon _{iA}\right) ^{-1}\left[ \left( \tilde{\varepsilon}%
_{i}^{r}-\varepsilon _{iA}\right) \overline{G_{i}^{<}}+\tilde{\varepsilon}%
_{i}^{<}\overline{G_{i}^{a}}\right] .  \label{Gdq}
\end{eqnarray}%
It can be easily verified that $x_{iA}\overline{G_{iA}^{r}}+x_{iB}\overline{%
G_{iB}^{r}}=\overline{G_{i}^{r}}$ and $x_{iA}\overline{G_{iA}^{<}}+x_{iB}%
\overline{G_{iB}^{<}}=\overline{G_{i}^{<}}$ as required by the physical
meaning of conditional Green's functions. In CPA-NVC theory, conditional
Green's functions were calculated with these expressions\cite{Ke 2008}.

It is shown in Appendix-\ref{app-B} that by solving $\overline{G_{iq}^{r}}$
and $\overline{G_{iq}^{<}}$ from NECPA Eqs.(\ref{NECPA Gr},\ref{NECPA Gd})
for binary disorder site $q=A,B$, the results are identical to Eqs.(\ref{Grq}%
,\ref{Gdq}). For disorder site with more than two impurity species, Eqs.(\ref%
{Grq},\ref{Gdq}) do not apply, and one has to rely on NECPA equations to
calculate conditional Green's functions.

\section{Numerical verification of NECPA}

Having established the NECPA theoretical framework, in this section we use
simple tight-binding (TB) models in one- and two-dimensions (1D, 2D) to
demonstrate the numerical accuracy of this theory.

\begin{figure*}[tbph]
\includegraphics[width=16cm, height=16cm]{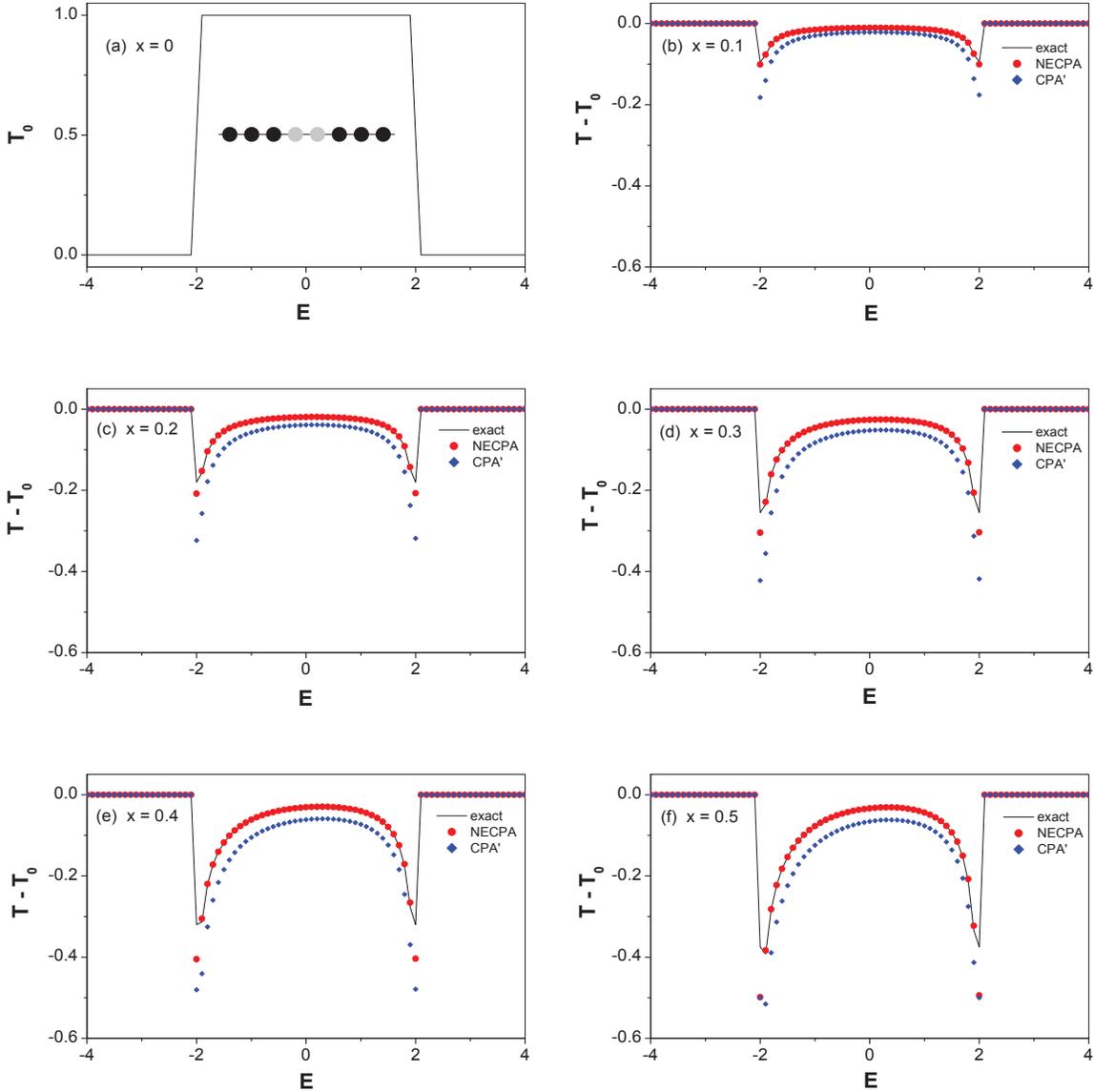} 
\caption{(color online) Transmission coefficient $T$ versus energy $E$ at
various disorder concentrations $x$ for 1D two-probe system. Three methods
are used for comparison: the exact solution, the NECPA method, and the CPA'
method that neglects the lesser coherent potential $\tilde{\protect%
\varepsilon}^{<}$ from the NECPA method. Each sub-figure is for a different
disorder concentration $x$. The inset of (a) shows the 1D TB model. Other
parameters are: $\protect\varepsilon _{0}=0$, $t=1$, $\protect\varepsilon %
_{1A}=\protect\varepsilon _{2A}=0$, $x_{1A}=x_{2A}=1-x$, $\protect%
\varepsilon _{1B}=\protect\varepsilon _{2B}=0.5$, $x_{1B}=x_{2B}=x$. }
\label{fig4}
\end{figure*}

\begin{figure*}[tbph]
\includegraphics[width=16cm, height=16cm]{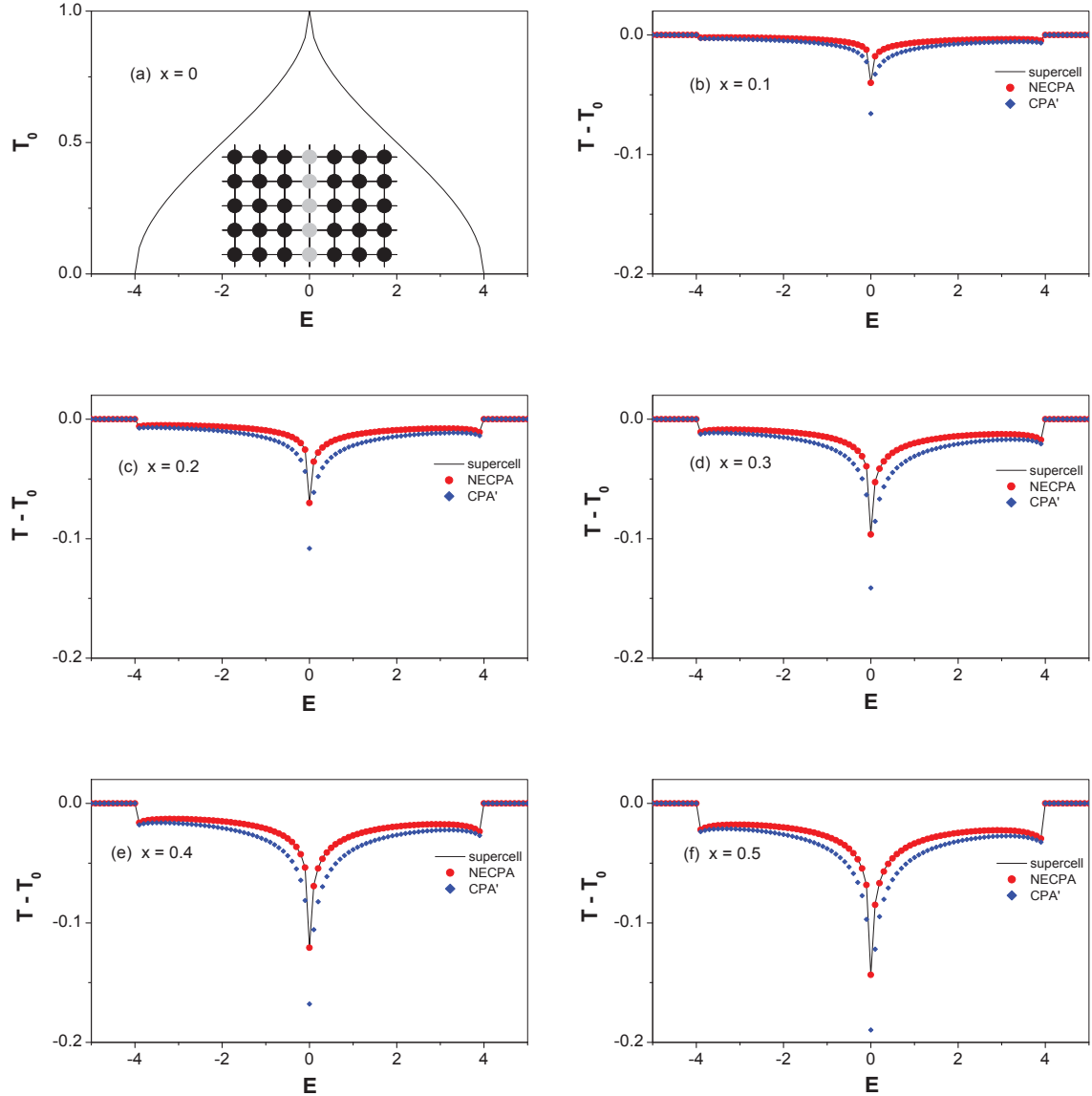} 
\caption{(color online) Transmission coefficient $T$ versus energy $E$ at
various disorder concentrations $x$ for 2D two-probe system. Three methods
are used for comparison: the supercell solution, the NECPA method and CPA'
method that neglects the lesser coherent potential $\tilde{\protect%
\varepsilon}^{<}$ from the NECPA method. Each sub-figure is for a different
disorder concentration $x$. The inset of (a) shows the 2D TB model which is
periodic in the transverse direction. Other parameters are: $\protect%
\varepsilon _{0}=0$, $t=1$, $\protect\varepsilon _{A}=0$, $x_{A}=1-x$, $%
\protect\varepsilon _{B}=0.5$, $x_{B}=x$. }
\label{fig5}
\end{figure*}

\subsection{One dimensional two-probe system}

\label{TB1D-subsec}

In this subsection, NECPA theory\ is applied to study a 1D TB model which
extends from $z=-\infty $ to $z=+\infty $ and contains two scatterers in the
central region, as shown in the inset of Fig.\ref{fig4}a. The black dots
represent clean sites having on-site energy $\varepsilon _{0}$; the gray
dots represent the disorder sites having on-site energy $\varepsilon _{i}$
which is a discrete random variable taking values $\varepsilon _{iq}$ ($%
q=1,2,\cdots $) with probability $x_{iq}$. Only nearest neighbors have
interactions with a coupling strength $t$. Due to the simplicity of 1D TB
model, the disorder average can be done \emph{exactly} by brute force
enumeration of all possible disorder configurations. For comparison, we
shall calculate the transmission coefficient $T\left( E\right) $ both by
NECPA equations and by brute force enumeration.

For NECPA calculation, disorder averaged Green's functions are calculated by
using Eqs.(\ref{NECPA Gr},\ref{NECPA Gd}). For the 1D TB model, the NECPA
equations are drastically simplified and the formula is listed in Appendix-%
\ref{app-C}. For brute force enumeration, disorder averaged Green's function
are calculated directly from its definition, namely
\begin{equation}
\left\{
\begin{array}{c}
\overline{G^{r}}=\sum_{q_{1}}%
\sum_{q_{2}}x_{1q_{1}}x_{2q_{2}}G_{q_{1}q_{2}}^{r},\notag \\
\\
G_{q_{1}q_{2}}^{r}=\left[ E-\left(
\begin{array}{cc}
\varepsilon _{1q_{1}} & t \\
t & \varepsilon _{2q_{2}}%
\end{array}%
\right) -\left(
\begin{array}{cc}
\Sigma _{0}^{r} & 0 \\
0 & \Sigma _{0}^{r}%
\end{array}%
\right) \right] ^{-1},\notag \\
\\
\overline{G^{<}}=\sum_{q_{1}}%
\sum_{q_{2}}x_{1q_{1}}x_{2q_{2}}G_{q_{1}q_{2}}^{<},\notag \\
\\
G_{q_{1}q_{2}}^{<}=G_{q_{1}q_{2}}^{r}\left(
\begin{array}{cc}
if_{L}\Gamma _{0} & 0 \\
0 & if_{R}\Gamma _{0}%
\end{array}%
\right) G_{q_{1}q_{2}}^{a},%
\end{array}%
\right.
\end{equation}%
in which $f_{L,R}$ are Fermi functions of the left and right electrodes, $%
\Gamma _{0}=-2\text{Im}\Sigma _{0}^{r}$ is the linewidth function of the
electrode. $\Sigma _{0}^{r}$ is the retarded self-energy of the electrode
which can be evaluated analytically for the semi-infinite 1D chain,%
\begin{equation*}
\Sigma _{0}^{r}=\xi \left( \frac{E+i0^{+}-\varepsilon _{0}}{t}\right) t,
\end{equation*}%
where
\begin{equation}
\xi \left( z\right) =\frac{z-i\sqrt{4-z^{2}}}{2},  \notag
\end{equation}%
in which the branch of the square root is chosen as Re$\sqrt{z}>0$.

Transmission coefficient $T\left( E\right) $ can be calculated with the aid
of disorder averaged Green's functions. Fig.\ref{fig4} plots $T\left(
E\right) $ at various disorder concentrations $x$ calculated by three
methods: the brute force enumeration; NECPA by evaluating Eqs.(\ref{TB1d Gr},%
\ref{TB1d Gd}) in Appendix-\ref{app-C}; and a CPA' method which is identical
to NECPA but neglecting contributions of the lesser coherent potential $%
\tilde{\varepsilon}^{<}$. In Fig.\ref{fig4}a, $T\left( E\right) $ is plotted
in the clean limit $x=0$ which is an integer step-like curve coinciding with
the number of conducting channels. In Fig.\ref{fig4}b to Fig.\ref{fig4}f, $%
T\left( E\right) $ is plotted with the increase of $x$. For clarity and to
show the effects of disorder scattering, a background transmission at $x=0$, $%
T_{0}\equiv \left[ T\left( E\right) \right] _{x=0}$, has been subtracted
from $\left[ T\left( E\right) \right] _{x>0}$. Several observations are in
order: (i) Transmission is suppressed gradually with the increase of $x$
which is a clear effect due to disorder scattering. The suppression of $%
T\left( E\right) $ by disorder is more pronounced in the vicinity of the
band edge. (ii) Results from NECPA agree quite well with the exact results
in full ranges of the disorder concentration, providing a verification of
the NECPA formalism. (iii) The CPA' solution has a noticeable deviation from
the exact results, indicating the importance of the lesser coherent
potential.

\subsection{Two dimensional two-probe model}

\label{TB2D-subsec}

In this subsection, NECPA theory\ is applied to study a 2D TB model which is
periodic in the transverse direction and contains a single layer of
scatterers in the central region, as shown in the inset of Fig.\ref{fig5}a.
As in the 1D TB model, the black dots represent clean sites having on-site
energy $\varepsilon _{0}$; the gray dots represent the disorder sites having
on-site energy $\varepsilon _{i}$ which is a discrete random variable taking
values $\varepsilon _{iq}$ ($q=1,2,\cdots $) with probability $x_{iq}$. Only
nearest neighbors have interactions with a coupling strength $t$. For the 2D
TB model, exact enumeration becomes impossible since the layer of scatterers
contains infinite number of disorder sites. Alternatively the disorder
average can be done by Monte Carlo simulation in a supercell. For
comparison, we shall calculate the transmission coefficient $T\left(
E\right) $ both by NECPA equations and by brute force supercell simulation.

For NECPA calculation, disorder averaged Green's functions are calculated by
using Eqs.(\ref{NECPA Gr periodic},\ref{NECPA Gd periodic}). The 2D TB model
illustrates how NECPA equations are applied to two-probe systems with
periodicity in the transverse direction and the formula is listed in
Appendix-\ref{app-C}. For brute force supercell simulation, we construct a
supercell of the two-probe system (see inset of Fig.\ref{fig5}a) and
disorder sites inside the supercell are occupied randomly according to the
probability $x_{iq}$. As long as the supercell is sufficiently large in the
transverse direction, the physics of a supercell two-probe system mimics
that of an infinite periodic two-probe system. In our simulation, the
supercell contains $1000$ rows and transmission is averaged over $100$
randomly generated disorder configurations.

Transmission coefficient $T\left( E\right) $ can be calculated with the aid
of disorder averaged Green's functions. Fig.\ref{fig5} plots $T\left(
E\right) $ at various disorder concentrations $x$ calculated by three
methods: the brute force supercell simulation as described above; NECPA by
evaluating Eqs.(\ref{TB2d Gr},\ref{TB2d Gd}) in Appendix-\ref{app-C}; and a
CPA' method which is identical to NECPA but neglecting contributions of the
lesser coherent potential $\tilde{\varepsilon}^{<}$. In Fig.\ref{fig5}a, $%
T\left( E\right) $ is plotted in the clean limit $x=0$ which can be well
understood by the 2D band structure of the TB model (not shown). In Fig.\ref%
{fig5}b to Fig.\ref{fig5}f, $T\left( E\right) $ is plotted with the increase
of $x$. For clarity and to show effects of disorder scattering, a background
transmission at $x=0$, $T_{0}\equiv \left[ T\left( E\right) \right] _{x=0}$,
has been subtracted from $\left[ T\left( E\right) \right] _{x>0}$. Several
observations are in order: (i) Transmission is suppressed gradually with the
increase of $x$ due to disorder scattering and the suppression is more
pronounced near the energies where $T\left( E\right) $ changes rapidly. (ii)
Results from NECPA agree quite well with the supercell results in full
ranges of the disorder concentration, providing a verification of the NECPA
formalism. (iii) The CPA' solution has a noticeable deviation from the exact
results, indicating the importance of the lesser coherent potential.

\section{Application of NECPA to DFT}

\label{nanodsim-sec}

As discussed in the Introduction, all realistic device materials contain
some degree of random disorders including atomic defects, vacancies, surface
roughness, interface irregularities and dopants, etc.. The NECPA formalism
presented above provides exciting opportunities to investigate disorder
effects in nanoelectronic systems from atomistic first principles. Without
disorders, the NEGF-DFT first principles formalism has been widely applied
to analyze nonlinear and nonequilibrium quantum transport properties of
nanoelectronics\cite{Jeremy}. With disorders, NEGF-DFT formalism need to be
generalized to NECPA-DFT formalism which is the subject of this section.

Notice that NECPA is only applicable to systems with substitutional
disorders. To apply NECPA to DFT, one has to work with localized atomic
orbitals in which only on-site blocks are different between the host atom and
impurity atom. There are many kinds of atomic orbital methods that have been
proved to work well with CPA, for instance, LMTO\cite{Anderson,Skriver}
(linear muffin-tin orbital), KKR\cite{Korringa,Kohn}
(Korringa-Kohn-Rostoker), FPLO\cite{FPLO} (full potential localized
orbital), etc.. It is straightforward to generalize CPA-DFT formalism of
those atomic orbital methods to NECPA-DFT formalism. For a concrete example,
the NECPA-LMTO method will be presented below in details.

LMTO is an atomic orbital implementation of DFT that has been widely applied
for decades in material physics to investigate electronic structures of
alloys, surfaces, and interfaces. For technical details of LMTO, we refer
interested readers to the monographs of Ref.%
\onlinecite{Anderson,Skriver,Turek}. In this section, we shall follow the
notation and terminology of Ref.\onlinecite{Turek}, and limit the LMTO
formalism to the minimum that is necessary for implementing NECPA.

Within LMTO there are two types of Green's functions: physical Green's
function and auxiliary Green's function. Physical Green's function is
directly related to physical quantities, while auxiliary Green's function is
the right one to apply NECPA. The relation between disorder averaged
physical Green's function ($\overline{G_{iq}^{r}}$ and $\overline{G_{iq}^{<}}
$) and auxiliary Green's function ($\overline{g_{iq}^{r}}$ and $\overline{%
g_{iq}^{<}}$) is as follows:
\begin{eqnarray}
&&\overline{G_{iq}^{r}}=\lambda _{iq}+\mu _{iq}\overline{g_{iq}^{r}}\mu
_{iq},  \notag \\
&&\overline{G_{iq}^{<}}=\mu _{iq}\overline{g_{iq}^{<}}\mu _{iq},\
\label{Gless3}
\end{eqnarray}%
where
\begin{equation}
\lambda _{iq}\equiv \frac{\gamma _{iq}-\alpha }{\Delta _{iq}+\left( \gamma
_{iq}-\alpha \right) \left( E-C_{iq}\right) },  \notag
\end{equation}%
\begin{equation}
\mu _{iq}\equiv \frac{\sqrt{\Delta _{iq}}}{\Delta _{iq}+\left( \gamma
_{iq}-\alpha \right) \left( E-C_{iq}\right) }.  \notag
\end{equation}%
In these expressions, $E$ is the energy, $\alpha $ is the screening
constant, $C_{iq}$, $\Delta _{iq}$, $\gamma _{iq}$ are potential parameters
of site-$i$ and component-$q$. Notice that in LMTO all the on-site variables
(e.g., $\overline{g_{iq}^{r}}$, $\overline{g_{iq}^{<}}$, $\overline{%
G_{iq}^{r}}$, $\overline{G_{iq}^{<}}$, $\lambda _{iq}$, $\mu _{iq}$) are $%
\left( l_{\max }+1\right) ^{2}$-by-$\left( l_{\max }+1\right) ^{2}$ matrices
in which $l_{\max }$ is the maximum orbital angular momentum quantum number
at the atomic site. Especially, $C_{iq}$, $\Delta _{iq}$, $\gamma _{iq}$, $%
\alpha $ are diagonal matrices of size $\left( l_{\max }+1\right) ^{2}$-by-$\left( l_{\max
}+1\right) ^{2}$. Since all the derivations of this work do not
assume that on-site variables are scalars, the formulation of NECPA remains
unchanged in LMTO.

To apply NECPA to auxiliary Green's function $\overline{g_{iq}^{r}}$ and $%
\overline{g_{iq}^{<}}$, one simply needs to carry out following replacement
in Eqs.(\ref{NECPA Gr periodic},\ref{NECPA Gd periodic}):%
\begin{equation}
\left\{
\begin{array}{c}
G\longrightarrow g \\
\\
H_{C}^{0}\left( k\right) \longrightarrow S\left( k\right) , \\
\\
E-\varepsilon _{iq}\longrightarrow P_{iq}\left( E\right) , \\
\\
E-\tilde{\varepsilon}^{r}\longrightarrow \tilde{P}^{r}\left( E\right) , \\
\\
-\tilde{\varepsilon}^{<}\longrightarrow \tilde{P}^{<}\left( E\right) .%
\end{array}%
\right.  \label{replacement}
\end{equation}%
The left hand side is the variable in general formulation and the right hand
side is the variable in LMTO language: $g$ is the auxiliary Green's
function; $S\left( k\right) $ is the Fourier transform of structure
constant; $P_{iq}\left( E\right) $ is the potential function defined as
\begin{equation*}
P_{iq}\left( E\right) \equiv \frac{E-C_{iq}}{\Delta _{iq}+\left( \gamma
_{iq}-\alpha \right) \left( E-C_{iq}\right) };
\end{equation*}%
$\tilde{P}^{r}$ and $\tilde{P}^{<}$ are retarded and lesser coherent
potentials%
\begin{eqnarray}
&&\tilde{P}^{r}\equiv diag\left( \left[ \tilde{P}_{1}^{r},\tilde{P}%
_{2}^{r},\cdots \right] \right) ,  \notag \\
&&\tilde{P}^{<}\equiv diag\left( \left[ \tilde{P}_{1}^{<},\tilde{P}%
_{2}^{<},\cdots \right] \right) .  \notag
\end{eqnarray}

Once $\overline{g_{iq}^{r}}$ and $\overline{g_{iq}^{<}}$ are solved from the
NECPA equations, $\overline{G_{iq}^{r}}$ and $\overline{G_{iq}^{<}}$ can be
calculated with Eq.(\ref{Gless3}). Consequently physical quantities can be
evaluated with these averaged Green's functions. The occupation number of
site-$i$ and specie-$q$ is obtained following Eq.(\ref{Ni1}),
\begin{equation}
N_{iq}=\text{Im}\int \frac{dE}{2\pi }\text{Tr }\overline{G_{iq}^{<}}\ \ .
\label{LMTO occ}
\end{equation}%
The density of states (DOS) is obtained by statistically weighted
contributions of DOS from each specie $q$,
\begin{equation}
D=-\frac{1}{\pi }\text{Im}\text{Tr}\sum_{iq}x_{iq}\overline{G_{iq}^{r}}.
\label{LMTO dos}
\end{equation}%
The transmission coefficient is obtained following Eq.(\ref{T2}),
\begin{equation}
T=\int_{-\pi }^{+\pi }\frac{dk}{2\pi }\text{Tr\ }\overline{g_{L}^{<}}\left(
k\right) \Gamma _{R}\left( k\right) ,  \label{LMTO trans}
\end{equation}%
in which $\overline{g_{L}^{<}}\left( k\right) $ is defined as $\overline{%
g^{<}}\left( k\right) $ with the substitution $f_{L}\rightarrow -i$ and $%
f_{R}\rightarrow 0$.

So far, given potential parameters $C_{iq}$, $\Delta _{iq}$, $\gamma _{iq}$,
disorder averaged Green's functions and physical quantities can be obtained
by solving NECPA equation. The complexity of NECPA-LMTO method comes from
the fact that potential parameters in turn depend on physical quantities and
are unknown \textit{a priori}. Therefore potential parameters must be solved
self-consistently together with disorder averaged Green's functions. The
flowchart for the self-consistent procedure of NECPA-LMTO method is
plotted in Appendix-\ref{flowchart}.

As an implementation of the NECPA-LMTO method, recently a new simulation
tool \emph{NanoDsim} has been developed\cite{NanoAcademic}. The software
aims at simulating nonequilibrium quantum transport properties of realistic
nanoelectronic devices from atomistic first principles. Some rather unique
features are worth mentioning, including: (i) it solves device Hamiltonian
of two-probe systems with atomic disorders at nonequilibrium
self-consistently within the general formalism of NECPA-DFT; (ii) it is
capable to simulate devices containing a few thousands atomic sites on a
moderate computer cluster\cite{IEEE review}; (iii) it has implemented a
recently proposed semi-local exchange-correlation potential\cite{MBJ XC}
thus providing good predictions of band gaps and dispersions for many common
semiconductors\cite{Yin TBP}; (iv) it has implemented a new post-analysis
tool, transmission fluctuation analyzer, to predict device variability due
to random discrete dopants \cite{Eric TBP}.

NanoDsim has been applied successfully to investigate nano-scaled devices.
In Ref.\onlinecite{Paul TBP}, the electronic potentials are simulated
atomically for Si nano-transistor channels with both n- and p-doping. The
results are in essentially perfect agreement with those obtained by
industrial TCAD software based on multitudes of material and electronic
input parameters. In Ref.\onlinecite{Jessie PRL}, realistic and important
device physics problems have been investigated providing useful microscopic
insights to improve device performance, namely how controlled localized
doping distributions in nanoscale Si transistors can suppress leakage
currents. In Ref.\onlinecite{Yin APL}, the band offset of the GaAs/Al$_{x}$Ga%
$_{1-x}$As heterojunctions is investigated for the entire range of the Al
doping concentration $0<x\leqslant 1$. The calculated band structures of the
GaAs, AlAs crystals and band gaps of the Al$_{x}$Ga$_{1-x}$As alloys, are in
very good agreement with the experimental results. We refer interested
readers to these literature for further details of NanoDsim simulations of
disorder effects in device physics.

\section{Summary}

In this work, we report the theory of NECPA for analyzing disorder effects in
nonequilibrium quantum transport. For the first time in literature, disorder
average is done within CPA on the complex-time contour, which provides a
\emph{unified derivation} of $\overline{G^{r}}$ and $\overline{G^{<}}$. To
accomplish the analytic continuation, the celebrated Langreth theorem is
generalized to include inverse operation. By applying the generalized
Langreth theorem to the contour ordered CPA equation, the NECPA equations, Eqs.(%
\ref{NECPA Gr}, \ref{NECPA Gd}), are derived for $\overline{G^{r}}$ and $%
\overline{G^{<}}$. In the low concentration limit, a set of analytical
solutions, Eqs.(\ref{lowX Gr}, \ref{lowX Gd}), have been obtained for NECPA
equations. For two-probe systems with transverse periodicity, NECPA
equations need to be adapted to include $k$-sampling in the transverse
dimensions, as derived in Eqs.(\ref{NECPA Gr periodic},\ref{NECPA Gd
periodic}).

Although the NECPA theory is mathematically equivalent to the CPA-NVC theory
developed previously, it has several advantages: (i) NECPA is elegant from a 
theoretical point of view due to its simplicity and unification; (ii) the
conditional Green's functions $\overline{G_{iq}^{r}}$ and $\overline{%
G_{iq}^{<}}$ are derived for disorder sites beyond the binary
situation; (iii) stable iterative solution methods are available for solving
NECPA equations.

The accuracy of the NECPA equations has been numerically verified by comparing to
brute force calculations of TB models. It is also demonstrated that NECPA
can be combined with the DFT technique to enable atomistic first principles
simulation of quantum transport. A simulation tool, NanoDsim, has been
developed as an implementation of NECPA-LMTO method. The software has
already been applied to a number of important and interesting device physics
problems.

Finally we would like to mention that NECPA equations are not limited to
combine with DFT. It is straightforward to apply NECPA to other atomistic
device models such as tight-binding models. In addition, the generalized
Langreth theorem, Eqs.(\ref{Langreth r},\ref{Langreth d},\ref{Langreth r3},%
\ref{Langreth d3},\ref{Langreth r2},\ref{Langreth d2}), can be applied to
other NEGF techniques (e.g, equation of motion) so that $G^{r}$ and $G^{<}$
are derived in a unified and consistent way.

\section*{ACKNOWLEDGEMENT}

We thank Youqi Ke for many discussions concerning the CPA-NVC theory and
its LMTO implementation. We thank Ke Xia for useful discussions on
LMTO method of DFT. We thank Yibin Hu for many discussions on the 
computational issues of the NECPA-LMTO and the NanoDsim software. We thank
Jesse Maassen, Ferdows Zahid, Yin Wang and Paul Zhang for
many discussions on practical applications of the NECPA-LMTO and the
NanoDsim software for first principles quantum transport simulations. The
financial support from NRC-IRAP of Canada is gratefully acknowledged.

\appendix

\section{Derivation of Eq.(\ref{CPA_conditional}) and Eq.(\ref{CPA_Omega})}

\label{maths}

In this appendix we derive Eq.(\ref{CPA_conditional}) and Eq.(\ref{CPA_Omega}%
) by using the contour ordered CPA equation (\ref{CPA_tao}) and the definition
of the conditional Green's function.

Firstly, we prove a lemma for the block matrix inverse. Assume that $A$ and $%
A^{\prime }$ are the inverse of following block matrices which are composed
of $2\times 2$ matrix blocks:%
\begin{eqnarray*}
A &=&\left(
\begin{array}{cc}
a_{11} & a_{12} \\
a_{21} & a_{22}%
\end{array}%
\right) ^{-1}, \\
A^{\prime } &=&\left(
\begin{array}{cc}
a_{11}^{\prime } & a_{12} \\
a_{21} & a_{22}%
\end{array}%
\right) ^{-1}.
\end{eqnarray*}%
It is straightforward to obtain%
\begin{eqnarray*}
A_{11} &=&\left( a_{11}-a_{12}a_{22}^{-1}a_{21}\right) ^{-1}, \\
A_{11}^{\prime } &=&\left( a_{11}^{\prime }-a_{12}a_{22}^{-1}a_{21}\right)
^{-1}.
\end{eqnarray*}%
It follows%
\begin{equation}
\left( A_{11}\right) ^{-1}-\left( A_{11}^{\prime }\right)
^{-1}=a_{11}-a_{11}^{\prime },  \label{lemma2}
\end{equation}%
which is the conclusion of the lemma.

Secondly, we apply the lemma to $\overline{G_{i}}$ and $\overline{G_{iq}}$
and obtain a useful relation between them. $\overline{G_{i}}$ and $\overline{%
G_{iq}}$ are defined in the second and third lines of Eq.(\ref%
{CPA_conditional}). Let $A=\overline{G_{i}}$ and $A^{\prime }=\overline{%
G_{iq}}$, and reorder $\overline{G_{i}}$ and $\overline{G_{iq}}$ such that
the block of site-$i$ is in the location of $a_{11}$ and $a_{11}^{\prime }$,
respectively. Due to the definition of $\tilde{\varepsilon}$ and $\tilde{%
\varepsilon}^{iq}$,
\begin{equation*}
a_{11}-a_{11}^{\prime }=\left( -\tilde{\varepsilon}\right) _{ii}-\left( -%
\tilde{\varepsilon}^{iq}\right) _{ii}=\left( -\tilde{\varepsilon}_{i}\right)
-\left( -\varepsilon _{iq}\right) .
\end{equation*}%
By using the lemma, it is derived%
\begin{equation}
\left( \overline{G_{i}}\right) ^{-1}-\left( \overline{G_{iq}}\right)
^{-1}=\varepsilon _{iq}-\tilde{\varepsilon}_{i}.  \label{xx1}
\end{equation}

Thirdly, we derive the first line of Eq.(\ref{CPA_conditional}). Substitute
Eq.(\ref{xx1}) into the second line of Eq.(\ref{CPA_tao}) and obtain%
\begin{equation}
t_{iq}=\left\{ \left[ \left( \overline{G_{i}}\right) ^{-1}-\left( \overline{%
G_{iq}}\right) ^{-1}\right] ^{-1}-\overline{G_{i}}\right\} ^{-1}.
\label{xx2}
\end{equation}%
Substitute Eq.(\ref{xx2}) into the first line of Eq.(\ref{CPA_tao}) and
obtain%
\begin{equation*}
\sum_{q}x_{iq}\left\{ \left[ \left( \overline{G_{i}}\right) ^{-1}-\left(
\overline{G_{iq}}\right) ^{-1}\right] ^{-1}-\overline{G_{i}}\right\} ^{-1}=0.
\end{equation*}%
By using $\sum_{q}x_{iq}=1$, Eq.(\ref{xx2}) can be simplified as
\begin{equation}
\sum_{q}x_{iq}\overline{G_{iq}}=\overline{G_{i}},  \label{xx3}
\end{equation}%
which is the first line of Eq.(\ref{CPA_conditional}).

Finally, we derive the fourth and fifth lines of Eq.(\ref{CPA_Omega}) from
Eq.(\ref{CPA_conditional}). As shown above, Eq.(\ref{CPA_conditional}) leads
to Eq.(\ref{xx1}), due to which one can always find a proper $\Omega _{i}$
such that%
\begin{equation}
\left( \overline{G_{i}}\right) ^{-1}=E-\tilde{\varepsilon}_{i}-\Omega _{i},
\label{xx4}
\end{equation}%
\begin{equation}
\left( \overline{G_{iq}}\right) ^{-1}=E-\varepsilon _{iq}-\Omega _{i}.
\label{xx5}
\end{equation}%
which are equivalent to the fourth and fifth lines of Eq.(\ref{CPA_Omega}).

\section{NECPA with principal layer algorithm}

\label{PL-subsec}

\begin{figure}[tbp]
\includegraphics[width=2in,angle=90]{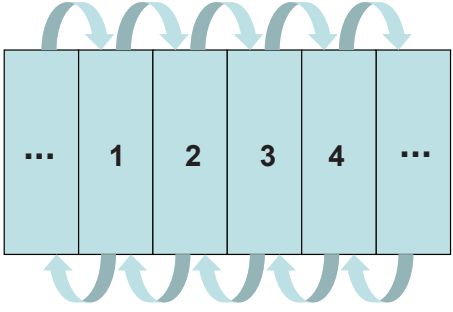}
\caption{(color online) Schematic plot showing the partition of a two-probe
system into principal layers along the transport direction. The integers
label the PLs. Partition is done in such a way that each PL only couples to
its nearest neighbor PLs.}
\label{fig3}
\end{figure}

In most two-probe systems, the size along the transport dimension is much
larger than their transverse dimensions. A two-probe system can therefore be
partitioned into principal layers (PL) along the transport direction such
that each PL only interacts with its two nearest neighbor PLs, as shown in
Fig.\ref{fig3}. Once a two-probe system is partitioned into PLs, the
calculation of $\overline{G^{r}}$ and $\overline{G^{<}}$ can be reduced to $%
\overline{G^{r}}=H^{-1}$ and $\overline{G^{<}}\equiv \frac{1}{i}%
H^{-1}D\left( H^{\dagger }\right) ^{-1}$. Here $H\equiv E-H_{C}^{0}-\tilde{%
\varepsilon}^{r}-\Sigma ^{r}$ is a block-tridiagonal matrix which has the
following non-zero pattern (illustrated with $N=4$ PLs):
\begin{equation}
H=%
\begin{tabular}{|c|c|c|c|}
\hline
$H_{11}$ & $H_{12}$ & $O$ & $O$ \\ \hline
$H_{21}$ & $H_{22}$ & $H_{23}$ & $O$ \\ \hline
$O$ & $H_{32}$ & $H_{33}$ & $H_{34}$ \\ \hline
$O$ & $O$ & $H_{43}$ & $H_{44}$ \\ \hline
\end{tabular}%
,  \label{H3}
\end{equation}%
where $H_{ij}$ are non-zero blocks, $O$ the zero blocks, and subscripts $i,j$
label the PLs. $D\equiv i\left( \Sigma ^{<}+\tilde{\varepsilon}^{<}\right) $
is a Hermitian block-diagonal matrix which has the following non-zero
pattern (illustrated with $N=4$ PLs):%
\begin{equation}
D=%
\begin{tabular}{|c|c|c|c|}
\hline
$D_{1}$ & $O$ & $O$ & $O$ \\ \hline
$O$ & $D_{2}$ & $O$ & $O$ \\ \hline
$O$ & $O$ & $D_{3}$ & $O$ \\ \hline
$O$ & $O$ & $O$ & $D_{4}$ \\ \hline
\end{tabular}%
,  \label{H4}
\end{equation}%
where $D_{i}=D_{i}^{\dagger }$ and subscript $i$ labels the PLs.

The calculation of $\overline{G^{r}}$ and $\overline{G^{<}}$ in the NECPA
equations involves matrix inversion and multiplication of these sparse
matrices. The computational cost of full matrix inversion and multiplication
is of $O\left( N^{3}\right) $ which becomes the bottleneck for large $N$.
Fortunately, only the diagonal elements of the Green's functions are needed
in solving NECPA equations. By taking the advantage of the zero blocks in
Eqs.(\ref{H3},\ref{H4}), the cost of calculating the diagonal elements can
be reduced to $O\left( N\right) $\cite{PrincipalLayer, PrincipalLayer2}.

Following Ref.\onlinecite{PrincipalLayer}, the calculation of diagonal
elements of $\overline{G^{r}}$ can be accomplished with the following
recursive relation:

\[
\overline{G_{ii}^{r}}\equiv A_{i}
\]
\begin{eqnarray}
C_{1} &=&H_{11}^{-1},  \nonumber \\
C_{i+1} &=&\left( H_{i+1,i+1}-H_{i+1,i}C_{i}H_{i,i+1}\right) ^{-1},
\nonumber
\end{eqnarray}%
\begin{eqnarray}
A_{N} &=&C_{N},  \nonumber \\
A_{i} &=&C_{i}+C_{i}H_{i,i+1}A_{i+1}H_{i+1,i}C_{i}.  \nonumber
\end{eqnarray}%
Following Ref.\onlinecite{PrincipalLayer2}, the calculation of diagonal
blocks of $\overline{G^{<}}$ can be accomplished with the following
recursive relation:%
\[
\overline{G_{ii}^{<}}\equiv \frac{1}{i}B_{i}
\]
\begin{eqnarray}
Y_{1} &=&D_{1},  \nonumber \\
Y_{i} &=&D_{i}+H_{i,i-1}C_{i-1}Y_{i-1}C_{i-1}^{\dagger }H_{i,i-1}^{\dagger },
\nonumber
\end{eqnarray}%
\begin{eqnarray}
B_{N} &=&C_{N}Y_{N}C_{N}^{\dagger },  \nonumber \\
B_{i} &=&C_{i}H_{i,i+1}B_{i+1}H_{i,i+1}^{\dagger }C_{i}^{\dagger
}-C_{i}Y_{i}C_{i}^{\dagger }+  \nonumber \\
&&A_{i}Y_{i}C_{i}^{\dagger }+\left( A_{i}Y_{i}C_{i}^{\dagger }\right)
^{\dagger }.  \nonumber
\end{eqnarray}%
The principal layer algorithm helps to optimize the performance of iterative
methods for solving NECPA equations.

\section{NECPA and CPA-NVC: the equivalence}

\label{app-A}

In this appendix we prove that $\overline{G^{<}}$ solved from NVC Eqs.(\ref%
{NVC 1},\ref{NVC 2}) are identical to $\overline{G^{<}}$ solved from NECPA
Eq.(\ref{NECPA Gd}) under the condition of NECPA Eq.(\ref{NECPA Gr}) for $%
\overline{G^{r}}$. Since the only difference between the expressions of $%
\overline{G^{<}}$ in NECPA Eq.(\ref{NECPA Gd}) and NVC Eq.(\ref{NVC 1}) are $%
\tilde{\varepsilon}^{<}$ and $\Lambda $, we proceed to prove that these two
quantities are actually identical. The NVC equation for $\Lambda $, Eq.(\ref%
{NVC 2}), is a nonhomogeneous linear equation which has a unique solution.
Hence the equivalence is proved if $\tilde{\varepsilon}^{<}$ satisfies Eq.(%
\ref{NVC 2}). It is shown below that $\tilde{\varepsilon}^{<}$ obtained from
NECPA Eq.(\ref{NECPA Gd}) indeed satisfies Eq.(\ref{NVC 2}).

We start by eliminating $\Omega _{i}^{<}$ from Eq.(\ref{NECPA Gd}):
\begin{eqnarray}
&&\overline{G_{i}^{<}}=\sum_{q}x_{iq}\overline{G_{iq}^{<}}=\sum_{q}x_{iq}%
\overline{G_{iq}^{r}}\Omega _{i}^{<}\overline{G_{iq}^{a}}  \notag \\
&=&\sum_{q}x_{iq}\overline{G_{iq}^{r}}\left[ \left( \overline{G_{i}^{r}}%
\right) ^{-1}\overline{G_{i}^{<}}\left( \overline{G_{i}^{a}}\right) ^{-1}-%
\tilde{\varepsilon}_{i}^{<}\right] \overline{G_{iq}^{a}}.  \label{A1}
\end{eqnarray}%
Next, we eliminate quantities $E$, $\varepsilon _{iq}$ and $\tilde{%
\varepsilon}_{i}^{r}$ from the expressions of $\overline{G_{iq}^{r}}$, $%
\overline{G_{i}^{r}}$ and $t_{iq}^{r}$ (see Eq.(\ref{NECPA Gr}) and Eq.(\ref%
{CPA_r}))
\begin{eqnarray*}
\overline{G_{iq}^{r}} &=&\left[ E-\varepsilon _{iq}-\Omega _{i}^{r}\right]
^{-1}, \\
\overline{G_{i}^{r}} &=&\left[ E-\tilde{\varepsilon}_{i}^{r}-\Omega _{i}^{r}%
\right] ^{-1}, \\
t_{iq}^{r} &\equiv &\left[ \left( \varepsilon _{iq}-\tilde{\varepsilon}%
_{i}^{r}\right) ^{-1}-\overline{G_{i}^{r}}\right] ^{-1},
\end{eqnarray*}%
and obtain
\begin{equation}
\overline{G_{iq}^{r}}=\overline{G_{i}^{r}}\left( 1+t_{iq}^{r}\overline{%
G_{i}^{r}}\right) .  \label{A2}
\end{equation}%
Finally, inserting Eq.(\ref{A2}) into Eq.(\ref{A1}) and using the CPA
condition of the first line of Eq.(\ref{CPA_r})
\begin{equation*}
\sum_{q}x_{iq}t_{iq}^{r}=\sum_{q}x_{iq}t_{iq}^{a}=0,
\end{equation*}%
we derive an equation for the quantity $\tilde{\varepsilon}_{i}^{<}$:
\begin{eqnarray}
\tilde{\varepsilon}_{i}^{<} &=&\sum_{q}x_{iq}t_{iq}^{r}\overline{G_{i}^{<}}%
t_{iq}^{a}-\sum_{q}x_{iq}t_{iq}^{r}\overline{G_{i}^{r}}\tilde{\varepsilon}%
_{i}^{<}\overline{G_{i}^{a}}t_{iq}^{a}  \notag \\
&=&\sum_{q}x_{iq}t_{iq}^{r}\left[ \overline{G^{r}}\left( \Sigma ^{<}+\tilde{%
\varepsilon}^{<}\right) \overline{G^{a}}\right] _{ii}t_{iq}^{a}-  \notag \\
&&\sum_{q}x_{iq}t_{iq}^{r}\overline{G_{i}^{r}}\tilde{\varepsilon}_{i}^{<}%
\overline{G_{i}^{a}}t_{iq}^{a},  \label{A3}
\end{eqnarray}%
which is identical to Eq.(\ref{NVC 2}).

\section{NECPA for binary disorder system}

\label{app-B}

In this appendix we show that for binary systems $q=A,B$, the conditional
Green's functions solved from NECPA Eqs.(\ref{NECPA Gr},\ref{NECPA Gd}) are
precisely given by Eqs.(\ref{Grq}, \ref{Gdq}) which are obtained from the
random variable method discussed in Sec. \ref{binary-subsec}.

First, we focus on retarded conditional Green's functions $\overline{%
G_{iA}^{r}}$ and $\overline{G_{iB}^{r}}$. Let $\overline{G_{iA}^{r}}=\lambda
_{iA}\overline{G_{i}^{r}}$ and $\overline{G_{iB}^{r}}=\lambda _{iB}\overline{%
G_{i}^{r}}$, we proceed to solve for the two coefficients $\lambda _{iA}$
and $\lambda _{iB}$. For the $q=A,B$ binary situation, NECPA equation (\ref%
{NECPA Gr}) is reduced to the following form
\begin{equation*}
\left\{
\begin{array}{c}
\overline{G_{i}^{r}}=x_{iA}\overline{G_{iA}^{r}}+x_{iB}\overline{G_{iB}^{r}},
\\
\\
\overline{G_{i}^{r}}=\left( E-\tilde{\varepsilon}_{i}^{r}-\Omega ^{r}\right)
^{-1}, \\
\\
\overline{G_{iA}^{r}}=\left( E-\varepsilon _{iA}-\Omega ^{r}\right) ^{-1},
\\
\\
\overline{G_{iB}^{r}}=\left( E-\varepsilon _{iB}-\Omega ^{r}\right) ^{-1}.%
\end{array}%
\right.
\end{equation*}%
By Eliminating quantities $\overline{G_{i}^{r}}$, $\overline{G_{iA}^{r}}$, $%
\overline{G_{iB}^{r}}$, and $E-\Omega ^{r}$ from the above, we obtain
\begin{eqnarray}
&&x_{iA}\lambda _{iA}+x_{iB}\lambda _{iB}=1,  \notag \\
&& \\
&&\left( \tilde{\varepsilon}_{i}^{r}-\varepsilon _{iA}\right) \left( \lambda
_{iA}^{-1}-1\right) ^{-1}=\left( \tilde{\varepsilon}_{i}^{r}-\varepsilon
_{iB}\right) \left( \lambda _{iB}^{-1}-1\right) ^{-1}\ .  \notag
\end{eqnarray}%
By using the normalization of probability $x_{iA}+x_{iB}=1$, $\lambda _{iA}$
and $\lambda _{iB}$ can be solved after some algebra:
\begin{eqnarray}
\lambda _{iA} &=&\frac{1}{x_{iA}}\left( \varepsilon _{iA}-\varepsilon
_{iB}\right) ^{-1}\left( \tilde{\varepsilon}_{i}^{r}-\varepsilon
_{iB}\right) ,  \notag \\
\lambda _{iB} &=&\frac{1}{x_{iB}}\left( \varepsilon _{iB}-\varepsilon
_{iA}\right) ^{-1}\left( \tilde{\varepsilon}_{i}^{r}-\varepsilon
_{iA}\right) ,  \notag
\end{eqnarray}%
which are identical to Eq.(\ref{Grq}) obtained from the random variable
method.

Similarly, the lesser conditional Green's function $\overline{G_{iA}^{<}}$
and $\overline{G_{iB}^{<}}$ can be solved from NECPA Eq.(\ref{NECPA Gd}) and
proved to be identical to Eq.(\ref{Gdq}) which is obtained from the random
variable method. However the algebra turns out to be extremely tedious. An
alternative way is to play the same trick as in the derivation of NECPA
equations, namely we generalize Eq.(\ref{Grq}) to a contour ordered form by
simply removing the superscript $r$,
\begin{eqnarray}
\overline{G_{iA}} &=&\frac{1}{x_{iA}}\left( \varepsilon _{iA}-\varepsilon
_{iB}\right) ^{-1}\left( \tilde{\varepsilon}_{i}-\varepsilon _{iB}\right)
\overline{G_{i}},  \notag \\
\overline{G_{iB}} &=&\frac{1}{x_{iB}}\left( \varepsilon _{iB}-\varepsilon
_{iA}\right) ^{-1}\left( \tilde{\varepsilon}_{i}-\varepsilon _{iA}\right)
\overline{G_{i}}.  \notag
\end{eqnarray}%
Now we apply the generalized Langreth theorem to these two equations and
straightforwardly obtain Eq.(\ref{Gdq}).

\section{NECPA equations for the TB models}

\label{app-C}

In the appendix, we list the NECPA equations as applied to the TB models.
For the 1D TB model, NECPA Eqs.(\ref{NECPA Gr},\ref{NECPA Gd}) are reduced
to the following form,
\begin{equation}
\left\{
\begin{array}{c}
\overline{G_{i}^{r}}=\sum_{q}x_{iq}G_{iq}^{r}, \\
\\
\overline{G^{r}}=\left[ E-\left(
\begin{array}{cc}
\tilde{\varepsilon}_{1}^{r} & t \\
t & \tilde{\varepsilon}_{2}^{r}%
\end{array}%
\right) -\left(
\begin{array}{cc}
\Sigma _{0}^{r} & 0 \\
0 & \Sigma _{0}^{r}%
\end{array}%
\right) \right] ^{-1}, \\
\\
\overline{G_{i}^{r}}=\left[ \overline{G^{r}}\right] _{ii}, \\
\\
\overline{G_{i}^{r}}=\left( E-\tilde{\varepsilon}_{i}^{r}-\Omega
_{i}^{r}\right) ^{-1}, \\
\\
G_{iq}^{r}=\left( E-\varepsilon _{iq}-\Omega _{i}^{r}\right) ^{-1}.%
\end{array}%
\right.  \label{TB1d Gr}
\end{equation}

\begin{equation}
\left\{
\begin{array}{c}
\overline{G_{i}^{<}}=\sum_{q}x_{iq}G_{iq}^{<}, \\
\\
\overline{G^{<}}=\overline{G^{r}}\left[ \left(
\begin{array}{cc}
\tilde{\varepsilon}_{1}^{<} & 0 \\
0 & \tilde{\varepsilon}_{2}^{<}%
\end{array}%
\right) +\left(
\begin{array}{cc}
if_{L}\Gamma _{0} & 0 \\
0 & if_{R}\Gamma _{0}%
\end{array}%
\right) \right] \overline{G^{a}}, \\
\\
\overline{G_{i}^{<}}=\left[ \overline{G^{<}}\right] _{ii}, \\
\\
\overline{G_{i}^{<}}=\overline{G_{i}^{r}}\left( \tilde{\varepsilon}%
_{i}^{<}+\Omega _{i}^{<}\right) \overline{G_{i}^{a}}, \\
\\
G_{iq}^{<}=G_{iq}^{r}\Omega _{i}^{<}G_{iq}^{a},%
\end{array}%
\right.  \label{TB1d Gd}
\end{equation}%
in which $\Gamma _{0}$ and $\Sigma _{0}^{r}$ have been defined in Sec. 
\ref{TB1D-subsec}. Eqs.(\ref{TB1d Gr},\ref{TB1d Gd}) are used in the NECPA
calculation of the 1D TB model presented in Sec. \ref{TB1D-subsec}.

The 2D two-probe model in Sec. \ref{TB2D-subsec} has transverse
periodicity, hence NECPA Eqs.(\ref{NECPA Gr periodic},\ref{NECPA Gd periodic}%
) are applied. For the 2D TB\ model, they are reduced to the following form,
\begin{equation}
\left\{
\begin{array}{c}
\overline{G^{r}}=\sum_{q}x_{q}G_{q}^{r}, \\
\\
\overline{G^{r}}=\int_{-\pi }^{+\pi }\frac{dk}{2\pi }\left[ E-2t\cos k-%
\tilde{\varepsilon}^{r}-2\Sigma _{0}^{r}\left( k\right) \right] ^{-1}, \\
\\
\overline{G^{r}}=\left( E-\tilde{\varepsilon}^{r}-\Omega ^{r}\right) ^{-1},
\\
\\
G_{q}^{r}=\left( E-\varepsilon _{q}-\Omega ^{r}\right) ^{-1}.%
\end{array}%
\right.  \label{TB2d Gr}
\end{equation}

\begin{equation}
\left\{
\begin{array}{c}
\overline{G^{<}}=\sum_{q}x_{q}G_{q}^{<}, \\
\\
\overline{G^{<}}=\int_{-\pi }^{+\pi }\frac{dk}{2\pi }\frac{if_{L}\Gamma
_{0}\left( k\right) +if_{R}\Gamma _{0}\left( k\right) }{\left\vert E-2t\cos
k-\tilde{\varepsilon}^{r}-2\Sigma _{0}^{r}\left( k\right) \right\vert ^{2}},
\\
\\
\overline{G^{<}}=\overline{G^{r}}\left( \tilde{\varepsilon}^{<}+\Omega
^{<}\right) \overline{G^{a}}, \\
\\
G_{q}^{<}=G_{q}^{r}\Omega ^{<}G_{q}^{a},%
\end{array}%
\right.  \label{TB2d Gd}
\end{equation}%
in which $\Gamma _{0}\left( k\right) =-2\text{Im}\Sigma _{0}^{r}\left(
k\right) $ and $\Sigma _{0}^{r}\left( k\right) $ is the Fourier transform of
the electrode self-energy:
\begin{equation}
\Sigma _{0}^{r}\left( k\right) =\xi \left( \frac{E+i0^{+}-\varepsilon
_{0}-2t\cos k}{t}\right) t.  \notag
\end{equation}%
The expression of $\xi (z)$ can be found in Sec. \ref{TB1D-subsec}.
Eqs.(\ref{TB2d Gr},\ref{TB2d Gd}) are used in the NECPA calculation of the
2D TB model presented in subsection \ref{TB2D-subsec}.

\section{Flowchart of NECPA-LMTO method}

\label{flowchart}

\begin{figure}[tbph]
\vspace{-1cm}
\includegraphics[height=9cm,width=11cm,angle=90]{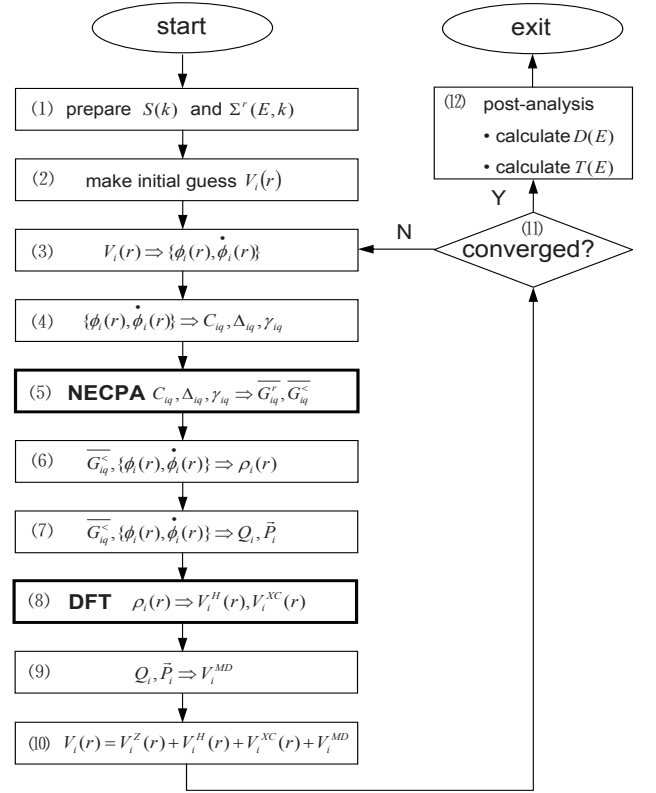}
\vspace{-0.5cm}
\caption{Flowchart of the NECPA-LMTO self-consistent procedure that is
implemented in the NanoDsim quantum transport package. For clarity, the
steps of NECPA and DFT have been highlighted.}
\label{fig6}
\end{figure}

The flowchart for the self-consistent procedure of NECPA-LMTO method is
sketched in Fig.\ref{fig6} and explained as follows.

\begin{enumerate}
\item Prepare the structure constant $S\left( k\right) $ and the lead
self-energies $\Sigma _{\beta }^{r}\left( E,k\right) $.

\item Make an initial guess of the atomic potential $V_{i}\left( r\right) $
for site-$i$.

\item Calculate orbitals $\left\{ \phi _{i}\left( r\right) ,\dot{\phi}%
_{i}\left( r\right) \right\} $ by solving Schr\"{o}dinger equation in the
potential $V_{i}\left( r\right) $.

\item Calculate potential parameters $C_{iq}$, $\Delta _{iq}$, $\gamma _{iq}$
with Wronskians involving $\phi _{i}\left( r\right) $ and $\dot{\phi}%
_{i}\left( r\right) $.

\item \textbf{NECPA}: Obtain disorder averaged Green's functions $\overline{%
G_{iq}^{r}}$, $\overline{G_{iq}^{<}}$ by solving NECPA equation containing $%
C_{iq}$, $\Delta _{iq}$, $\gamma _{iq}$. See Eqs.(\ref{Gless3},\ref{NECPA Gr
periodic},\ref{NECPA Gd periodic},\ref{replacement}).

\item Calculate the atomic charge density $\rho _{i}\left( r\right) $ by
using $\overline{G_{iq}^{<}}$ and $\left\{ \phi _{i}\left( r\right) ,\dot{%
\phi}_{i}\left( r\right) \right\} $.

\item Calculate the monopole $Q_{i}$ and the dipole $\overrightarrow{P_{i}}$
by using $\overline{G_{iq}^{<}}$ and $\left\{ \phi _{i}\left( r\right) ,\dot{%
\phi}_{i}\left( r\right) \right\} $.

\item \textbf{DFT}: Calculate the Hartree potential $V_{i}^{H}\left(
r\right) $ and the exchange-correlation potential $V_{i}^{XC}\left( r\right)
$ by using $\rho _{i}\left( r\right) $.

\item Calculate the Madelung potential $V_{i}^{MD}$ by using $\left\{ Q_{i},%
\overrightarrow{P_{i}}\right\} $ with Ewald summation.

\item Update the atomic potential $V_{i}\left( r\right) =V_{i}^{Z}\left(
r\right) +V_{i}^{H}\left( r\right) +V_{i}^{XC}\left( r\right) +V_{i}^{MD}$
where $V_{i}^{Z}\left( r\right) \equiv -\frac{Z}{r}$ is the nuclear
potential.

\item Check the convergence of $V_{i}\left( r\right) $. Go back to step-3
until $V_{i}\left( r\right) $ is fully converged for every atomic site.

\item Carry out post-analysis: calculate density of states by using Eq.(\ref%
{LMTO dos}); calculate transmission coefficient by using Eq.(\ref{LMTO trans}%
), etc..
\end{enumerate}

The above procedure involves technical details of LMTO method, we refer
interested reader to Ref.\onlinecite{Turek}{.}


\begin{thebibliography}{99}
\bibitem{comment0} For an atomic disorder site discussed in Section \ref%
{nanodsim-sec}, the on-site energy is a matrix block rather than a scalar.
\bibitem{comment1} If disorder sites appear everywhere in the electrodes
which extend to $z=\pm \infty $, traveling waves will not be able to
propagate and no current can flow.
\bibitem{comment3} From the Feynman diagrammatic point of view, both CPA and NVC neglect diagrams with crossing impurity lines and discard diagrams with dangling impurity lines. Since the diagram structures of CPA and NVC are different, it is not immediately transparent why the neglected terms are equal.
\end{thebibliography}
\end{document}